\documentclass[11pt,letterpaper]{article}
\pdfoutput=1

\usepackage{graphicx,array}
\usepackage[dvipsnames]{xcolor}
\definecolor{darkblue}{rgb}{0,0.1,0.5}
\definecolor{darkgreen}{rgb}{0,0.5,0.2}
\definecolor{seablue}{rgb}{0,0.2,0.6}
\usepackage{latexsym}
\usepackage{amssymb, amsmath}
\usepackage{bm}        
\usepackage[numbers,sort&compress]{natbib}
\usepackage{bm,relsize}
\usepackage{slashed}
\usepackage{tikzfeynman}
\usepackage{mathrsfs}
\usepackage{hyperref} 
\hypersetup{
    colorlinks=true,       
    linkcolor=darkblue,          
    citecolor=BrickRed,        
    filecolor=magenta,      
    urlcolor=MidnightBlue           
}
\usepackage[all]{hypcap} 
\usepackage{subcaption}

\usepackage[font=small,labelfont=bf,labelsep=period]{caption}

\newcommand{\redcross}{ \textcolor{red}{\otimes}}

\newcommand{\SE}{\mathrm{SE}}

\newcommand{\CG}{{\rm CG}}

\newcommand{\be}{\begin{equation}}
\newcommand{\ee}{\end{equation}}

\begin{document}

\begin{flushright}

\end{flushright}
\vspace{.6cm}
\begin{center}
{\LARGE \bf 
Cosmological Production of Dark Nuclei
}
\bigskip\vspace{1cm}{

\large Michele Redi and Andrea Tesi}
\\[7mm]
 {\it \small
INFN Sezione di Firenze, Via G. Sansone 1, I-50019 Sesto Fiorentino, Italy\\
Department of Physics and Astronomy, University of Florence, Italy
 }

\end{center}

\bigskip \bigskip \bigskip \bigskip

\centerline{\bf Abstract} 
\begin{quote}
We study the formation of Dark Matter nuclei in scenarios where DM particles are baryons of a new confining gauge force.
The dark nucleosynthesis is analogous to the formation of light elements in the SM and requires as a first step the formation of dark deuterium.
We compute this process from first principles, using the formalism of pion-less effective theory for nucleon-nucleon interactions.
This controlled effective field theory expansion allows us to systematically compute the cross sections for generic SM representations under the assumption of shallow bound states.
In the context of vector-like confinement models we find that, for nucleon masses in the TeV range, baryonic DM made of electro-weak constituents can form a significant fraction 
of dark deuterium and a much smaller fraction of dark tritium. Formation of dark nuclei can also lead to monochromatic photon lines in indirect detection.
Models with singlets do not undergo nucleosynthesis unless a dark photon is added to the theory.
\end{quote}

\vfill
\noindent\line(1,0){188}
{\scriptsize{ \\ E-mail:\texttt{ \href{mailto:michele.redi@fi.infn.it}{michele.redi@fi.infn.it}, \href{andrea.tesi@fi.infn.it}{andrea.tesi@fi.infn.it}}}}
\newpage

\tableofcontents

\setcounter{footnote}{0}

\section{Introduction}

The stability of Dark Matter (DM) on cosmological time scales strongly suggests the existence of new accidental symmetries in Nature.
In a minimalistic approach where DM is a representation of the Standard Model (SM) gauge group, the possibilities that DM  is accidentally stable are very limited and constrained \cite{Cirelli:2005uq}. A different class of models where DM is accidentally stable with a wide variety of SM quantum numbers is the one where DM is a baryon-like object of some strongly interacting dark sector \cite{Antipin:2015xia,Mitridate:2017oky,Kribs:2016cew}, in the framework of vector-like confinement \cite{Kilic:2009mi}. In these scenarios, the presence of elementary `dark' fermions with an accidental U(1) symmetry guarantees the stability of the lightest dark baryons,  as the stability of protons and nuclei is related to baryon number conservation in the SM.

In such scenarios, dark nuclear forces are expected to give rise to dark nuclei. For example in \cite{mccullough1,mccullough2} it was shown that bound states with baryon number 2 exist and the absence of a Coulomb barrier implies that states with  large baryon number very likely exist in the spectrum. The possibility that DM is confined into more complex structures is of obvious theoretical and experimental  interest.

In this work we wish to quantitatively study the nucleosynthesis of the dark sector. 
In the SM, the formation of light elements during Big Bang Nucleosynthesis (BBN) depends upon a few odd circumstances: for example on the fact that the deuterium binding energy is small and comparable to the proton-neutron mass difference. We find that even for baryonic DM, the success of nucleosynthesis  depends on the formation of dark deuterium 
but, contrary to the SM,  the different densities and binding energy of DM require a precise knowledge of the dark deuterium production cross-section.

First, we establish the general features of dark nucleons and nuclei in models with a new confining gauge interaction and fermions in vector-like representations. Under broad assumptions the spectrum is dominated by the nuclear binding energies $E_B$ much smaller than the nucleon mass $M_N$ and electro-weak effects can be included perturbatively. Importantly the synthesis of nuclei requires release of energy which automatically allowed in models with electro-weak constituents where it is always possible to radiate a photon,
\be\label{BSF}
\begin{tikzpicture}[line width=1.5 pt, scale=1.7]
	\node at (-1.2,0.25) {$N$};
	\node at (-1.2,-0.25) {$N$};
	\draw[] (-1,0.25)--(0,0);
	\draw[] (-1,-0.25)--(0,0);
	\draw[line width=4pt, color=red] (1,0)--(0,0);
	\draw[vector,color= blue] (0,0)--(0.7,.7);
	\draw[fill, color=gray] (0,0) circle (.3cm);
	\node at (-1.8,0) {$(\mathbf{r},S)\, \LARGE \bigg\{$};
	\node at (1.8,0) {$D\, (\mathbf{r}',S')$};
	\node at (1.05,.7) {$\ \\ ~~ \gamma, W, Z$};
\end{tikzpicture}
\nonumber\,
\ee

The key input to determine the abundance of nuclei is the deuterium cross-section. At first sight its calculation involves strongly coupled nuclear reactions that seem difficult to control. This is not the case, however, in light of the smallness of $E_B/M_N$ and a precise  computation is possible for shallow bound states. We will get inspiration from effective field theory of nucleon interactions \cite{Kaplan:1998tg}, applying the same techniques to the case of dark nucleons with arbitrary quantum numbers $(\mathbf{r},S)$.
The cross-section for bound state formation through electric (dipole) and magnetic interactions are found,
\begin{equation}
\sigma v_{\rm rel}\big|_{\rm electric} = K_E  \frac {2\pi \alpha^3}{v_{\rm rel}^3}\times \frac {\pi \alpha}{M_N^2}\left( \frac {M_N}{E_B}\right)^{\frac 1 2} v_{\rm rel}^2\,,\quad
\sigma v_{\rm rel}\big|_{\rm magnetic}= K_M  \frac {2\pi \alpha}{v_{\rm rel}}\times \frac {\pi \alpha}{M_N^2}\left( \frac {E_B}{M_N}\right)^{\frac 3 2}\,,
\nonumber
\end{equation}
where the first factor accounts for possible Sommerfeld enhancement (SE) and $K_E$ and $K_M$ are group theory factors. 
With these result in hand we can easily compute the abundance of deuterium and heavier nuclei in a given model by solving the relevant Boltzmann equations. For models with electro-weak charges we find that a fraction of DM  is bound in deuterium and heavier elements are not significantly produced. 
Formation of nuclei through photon emission can be tested in indirect detection experiments such as FERMI.

The present work clarifies how in general the first steps of dark nuclei formation can occur, based on first principle calculations. In this respect it provides a quantitative input for works that focus on the asymptotic fusion of large nuclei \cite{Hardy:2014mqa,Hardy:2015boa}, and also for models where the dark baryon coupled to dark photons as in \cite{1406.1171}, that we reconsider. On a more technical side, this paper extends the analysis of cosmological bound state formation of perturbative bound states \cite{vonHarling:2014kha,Mitridate:2017izz} to strongly coupled nuclei. 

The paper is organised as follows. After an introduction to the properties of nuclei in vector-like confinement scenarios in section \ref{sec:darkforce}, we derive in section \ref{sec:estimates}  the Boltzmann equations for dark deuterium in a few models. We compute the deuterium formation cross section with non-relativistic effective field theory techniques in section \ref{sec:calculation} and the appendix. Section \ref{sec:triplet}  discusses the case of dark nuclei made of SU(2)$_L$ triplets, while in section \ref{sec:darkphoton} we consider dark nuclei charged under a dark photon. We conclude and outline future directions in section \ref{sec:conclusions}.


\section{Dark Force}
\label{sec:darkforce}
We frame our discussion of DM nuclei within the scenarios of vector-like confinement \cite{Kilic:2009mi}.
The SM is extended with a new non abelian gauge force and fermions charged under the SM and dark group, described by the
renormalisable lagrangian,
\begin{equation}
\mathscr{L}=\mathscr{L}_{\rm SM}- \frac { {\cal G}_{\mu\nu}^a{\cal G}^{a\,\mu\nu}} {4 g_{\rm DC}^2}+{\bar \Psi}(i\slashed{D}-m_\Psi)\Psi\,.
\end{equation}
Such framework has special interest for DM as it automatically generates accidentally stable DM candidates \cite{Antipin:2015xia}.
We will focus  on SU(N)$_{\rm DC}$ gauge theories with fermions in the fundamental of dark colour\footnote{In models with real reps such as fundamental of SO($N$)  or adjoint of SU($N$) \cite{Contino:2018crt} baryon number is only conserved modulo 2 so they do not support stable nuclei with $A>1$. Models with pseudo-real representations such as fundamental of Sp($N$) do not have stable baryons.} and masses $m_\Psi$ smaller than the confinement scale $\Lambda$. 

Upon confinement the spectrum consists of dark pions and dark baryons with charges under the SM determined by their constituents.
The theory features an accidental U(1) symmetry, the dark baryon number, under which the fermions $\Psi$ transform with the same phase. This symmetry guarantees the stability of the lightest baryon, which for appropriate choices of SM representation can be a viable DM candidate. The very same dark baryon number also guarantees the exact stability of the lightest states in each charge sector. 
This leads to the stability or metastability of nuclei or more in general state of matter that carry baryon number. 
The quantum numbers under the SM are uniquely fixed as gauge interactions naturally select the smallest representation as the most bound.

As an example we will consider the simplest models with SU(3)$_{\rm DC}$ gauge group:
\begin{itemize}
\item $\Psi$ is a  triplet under a SU(2)$_L$. The lightest baryon $V\equiv\Psi\Psi\Psi$ is also a triplet and has spin 1/2. The lightest states are pions 
transforming as a triplet and quintuplet with mass splitting $\Delta M_\Pi^2\sim \alpha_2/(4\pi) \Lambda^2$. The triplet is accidentally stable due to G-parity \cite{hill},
but it can decay through dimension-5 operators.
\item $\Psi$ is a  singlet of the SM. In this context some heavier unspecified charge fermions are needed to guarantee thermal contact with the SM
but play no other role in the dynamics. The lightest baryon has spin 3/2 for one flavour of singlets or 1/2 for more flavours. 
\end{itemize}

Previous studies focussed on symmetric DM where the relic abundance is generated through thermal decoupling, which leads to a baryon mass
around 100 TeV to reproduce the known DM density. Here  will focus on asymmetric DM as this this maximises the formation of nuclei. 
This requires a suppressed  symmetric component so naturally $M_N < 100$ TeV. Direct searches at the LHC place a bound of around 1 TeV 
for the mass \cite{Barducci:2018yer,Kribs:2018ilo}, while the singlets just need to be heavier than about 10 MeV to avoid bound on number of species.

\subsection{Properties of Dark Nuclei}\label{sec:properties}

At zero temperature dark nucleons are expected to bind into larger nuclei due to residual strong interactions.
Based on nuclear physics examples, as well as lattice results \cite{Beane:2012vq}, we consider as typical binding energies,
\begin{equation}
0.001< E_B/M_N < 0.1\,.
\end{equation}
In the SM due to the electro-static repulsion, only nuclei with atomic number $A\lesssim 100$ are long lived or cosmologically stable. 
Moreover the presence of bottlenecks associated to the details of the nuclear spectrum implies that only the lightest nuclei are synthesised cosmologically. 

For accidental DM models with electro-weak constituents  the situation is likely different. 
If the mass of the dark baryon is larger than the electro-weak splittings, it is possible to exploit the approximate SU(2)$_L$ symmetry to classify nuclei into SM representations at least for the nuclei with small atomic number $A$. Actually, in the limit where we neglect SM interactions the theory has a larger accidental global symmetry SU($N_F$) corresponding to the dimension of the lightest nucleon representation. All the states can thus be classified into multiplets of SU($N_F$). Electro-weak interactions break this symmetry splitting the lightest nucleon multiplet into SM multiplets by an amount,
\begin{equation}
\Delta M_N \sim I_N(I_N+1) \frac {\alpha_2}{4\pi} M_N
\label{nucleonsplitting}
\end{equation}
where $I_N$ is the isospin of the nucleon.

The electro-weak splitting of nucleons induces a splitting between the dark nuclei. Since the shift above
can be larger than the nuclear binding energies, the splitting of nuclei made of different nucleon representation is dominated by the splitting of constituents over a large range of parameters.
One consequence is that as in the SM heavier baryons do not participate to nucleosynthesis. 
Cosmologically the bound state formation begins at temperatures of  order $E_B/20$, where $E_B$ is the binding energy.
Since the heavier nucleons decay into the lighter ones through strong interactions we can neglect the population of heavier nucleons as long as,
\begin{equation}
\frac {E_B} {20}< \Delta M_N 
\longrightarrow  \frac {E_B} {M_N}< \alpha_2
\end{equation}
This is the typical range expected for nuclear binding energies.

Thus we can focus on the nuclei made of the lightest SM rep. The nuclei made of the lightest nucleon multiplet can be decomposed into SU(2)$_L$ reps
split by nuclear binding energies $\Delta E_B^N$. The electro-weak correction to the binding energy is given by,
\begin{equation}
\Delta E_B^W \sim \lambda \frac{\alpha_2}{R_{\rm nucleus}}
\end{equation}
where $\lambda=I_N(I_N+1)-I_B(I_B+1)/2$ is the effective coupling in the isospin channel of the nucleus.
For shallow bound states we estimate the size of the nucleus with the scattering length $1/a=\sqrt{M_N E_B}$, see section \ref{sec:calculation}.
It follows that the nuclear binding energies dominate for $E_B/M_N> 10^{-3}$. Electro-weak corrections can  be included perturbatively in the relevant region
of parameters, and for $R_{\rm nucleus}< M_Z^{-1}$  the SM gauge fields can be treated as massless.

For large $A$ a multitude of SM reps exist with isospin up to $ A I$ where $I$ is the nucleon isospin. 
Of these the smaller representations are attractive making the nuclei more bound while the largest representation have a Coulombian
energy that scales as $A I (A I+1)$ that unbind nuclei of arbitrarily large charges.  In light of this, the valley of stability of dark nuclei will likely extend to very large to $A$, 
at least for small electro-weak charges.

Only the lightest baryon in each baryonic number sub-sector will be stable so that at late times all baryons produced cosmologically 
decay to a neutral state with isospin 0 or 1/2. This process is controlled by the decay rates among and within isospin multiplets induced by
electro-weak interactions, analogously to de-excitation of hydrogen atom. In particular,
\begin{itemize}
\item Each isospin multiplet can decay  to states with smaller isospin.
Since we focus on the $s$-wave bound states the rate is dominated by emission of a photon through magnetic dipoles with $\Delta S=1$. 
As we show in appendix \ref{sec:appB} the rate can be computed model independently as,
\begin{equation}
\Gamma \approx \alpha  \frac{(E_{B_\mathbf{1}}-E_{B_2})^2}{M_N^2} \sqrt{E_{B_\mathbf{1}} E_{B_2}}
\end{equation}
where $B_{1,2}$ are the binding energies of two bound states with $\Delta I=\Delta S=1$.

\item  The splitting within each baryonic electro-weak multiplet is given by \cite{Cirelli:2005uq},\footnote{The size of the nuclei scales as $A^{1/3}/M_N$, 
therefore they can be treated as elementary as long as the size is smaller than $1/M_W$.}
\begin{equation}
\Delta M_N= Q^2  \alpha_2 M_W \sin \frac{{\theta}_W}2
\end{equation}
which for an SU(2)$_L$ triplet gives $\Delta M_N=165$ MeV. Since charged and neutral components remain in equilibrium, the abundance of charged nuclei is then approximately $n_{V^+}\approx 2 n_{V^0}$ for $T> \Delta M$.
\end{itemize}

In what follows we will study the cosmological synthesis of dark nuclei. Cosmologically states with large $A$ could be produced through fusion processes via aggregation of heavy elements. In \cite{Hardy:2014mqa,1406.1171} it was argued that at least for light DM the dark synthesis is very efficient and one ends up with a distribution of large nuclear states. These studies overlook the first step of formation of nuclei with baryon number 2 (dark deuterium) that cannot take place through simple fusion processes but requires 
some energy to emitted. As we will show the deuterium abundance is often suppressed leading to a small abundance of larger nuclei.

\section{Deuterium Abundance}\label{sec:estimates}

As in the SM, we assume a separation of scales between the nucleon masses $M_N$ and the nuclear binding energies $E_B$. 
In this case the treatment of dark nucleosynthesis becomes relatively simple. 
At temperatures below $E_B$, when nuclear reactions can form nuclei, the nucleons are non-relativistic and already decoupled from the SM plasma. The yield of dark matter number is then set by
\be\label{YDM}
Y_{\rm DM}=\frac{n_{\rm DM}}{s}=4.3\, \times \, 10^{-13} \ \bigg(\frac{\mathrm{TeV}}{M_N}\bigg)\,,
\ee
where $n_{\rm DM}$ is the DM number density and $s=(2 \pi^2/45) g_\star T^3$ the entropy of the plasma and $g_\star$ the relativistic degrees of freedom of the plasma at a temperature $T$.

We will assume for simplicity that DM is asymmetric, since this maximises the production of bound states.
Given that thermal abundance indicates $M_N\sim 100$ TeV we assume the DM mass to be below this value.
This is also necessary to produce a significant fraction of nuclei.

Dark BBN takes place during a period where the dark baryon number is conserved, that is
\be\
1 = \sum_{A, \{i\}} \frac{A Y_{A_i}}{Y_{\rm DM}}\equiv\sum_{A, \{i\}}X_{A_i}\,,
\ee
where we have introduced the mass fractions $X_{A_i}=A Y_{A_i}/Y_{\rm DM}$ of a nucleus with atomic number $A_i$. 
Practically, we are only tracking how the total DM yield in eq.~(\ref{YDM}) gets redistributed among different nuclear species. 

A successful nucleosynthesis of states with $A\gg 2$ depends on the efficiency of deuterium formation, similarly to the SM case.
In this section, we derive the parametric dependence of the dark deuterium abundance on the relevant parameters of the theory with a focus on two scenarios
\begin{itemize}
\item SM charged constituents, with production $N+N\to D +X$, with $X=W,Z,\gamma$ .
\item SM neutral constituents, with production $N+N+N\to D+N$.
\end{itemize}

In general bound state formation requires some energy to be released.
We do not consider the possibility of emitting a pion of the strong sector. At first sight this process could be favoured 
by the strong coupling to nucleons but it is likely forbidden kinematically. Indeed, based on nuclear physics examples \cite{Beane:2012vq}, we expect the following scaling
\begin{equation}
\sqrt{M_N E_B} \sim 0.3 m_\pi\,.
\end{equation}
This is particularly significant for models with singlets where
no other light states exist.

\subsection{SM charged constituents}\label{subsec:SMcostituents}
When the constituents of DM have electro-weak charges, direct searches at LHC imply that the scale of new states must be larger than about 1 TeV \cite{Barducci:2018yer,Kribs:2018ilo}.\footnote{We will focus mostly on electro-weak constituents that are more naturally compatible with direct detection bounds. See however \cite{DeLuca:2018mzn} for counter-examples.}
For this value of the mass the numerical density in eq.(\ref{YDM}) is much smaller than the yield of baryonic matter, suggesting that the formation of bound states less likely than in the SM.

We consider the first step of formation of dark deuteron $N+N \to D+ X$, where $X$ stands for an electro-weak gauge boson $\gamma/Z/W$ in  equilibrium with the SM bath. For simplicity we assume that no other channels contribute, so that the Boltzmann equation for $D$ takes the form
\begin{equation}\label{eq:2to2}
\dot n_D+ 3H n_D= \langle \sigma_{D} v\rangle\left[ n_N^2- \frac {(n_N^{\rm eq})^2}{n_D^{\rm eq}}n_D\right]\,,
\end{equation}
where $n_B$ and $n_D$ are the numerical densities of dark baryons and dark deuterium and we assume radiation domination. During radiation domination, the Hubble parameter is 
\be
H= \sqrt{g_\star (T)\frac{\pi^2}{90}} \frac{T^2}{M_{\rm Pl}}\,,\quad M_{\rm Pl}\equiv2.4 \times 10^{18}\, \mathrm{GeV}\,.
\ee
The second term in the square brackets of eq.~\eqref{eq:2to2} becomes exponentially suppressed as $e^{-E_B/T}$, when the temperature drops below the binding energy. 
At this stage, dark deuterium can be formed, so that it is convenient to introduce a new time variable $z\equiv B/T$.
In terms of the mass fractions, the Boltzmann equation can be cast in the following form,
\begin{equation}\label{boltzmann22}
\frac {d X_{D}}{dz}= 2\frac{c   \sqrt{g_\star} M_{\rm Pl} E_B  \langle\sigma_{D} v\rangle}{z^2} Y_{\rm DM}\bigg[ (1-X_D)^2 - \beta \big(\frac{g_N^2}{g_D g_\star}\big)  \big(\frac{z^{3/2} e^{-z}}{Y_{\rm DM}}\big)\big( \frac{M_N}{E_B}\big)^{3/2} \frac{X_D}{2} \bigg]\,
\end{equation}
where $c=1.32$,  $\beta=45/(16 \pi^{7/2})$ and $g_*$ is the effective number of degrees of freedom. Only at a temperature significantly lower than the binding energy the bound state can be produced without being immediately dissociated. Given the smallness of $Y_{\rm DM}$, this happens at values of $z_f\approx 20$, when the coefficient of the term linear in $X_D$ inside the square brackets becomes of O(1). At this time we have the most efficient stage of deuterium synthesis starting with boundary condition $X_D(z_f)\approx 0$. Soon after, $X_D$ approaches a constant value that is linearly sensitive to the product of the binding energy and the cross-section, given by
\be
\frac{X_D}{1-X_D}=2 c\sqrt{g_\star} M_{\rm Pl} E_B    Y_{\rm DM} \int_{z_f}^\infty dz \frac{\langle \sigma_{D} v\rangle}{z^2}\,.
\ee
Away from saturation, $X_D\ll 1$, we have the following for an $s$-wave cross section,
\be
X_D = 5\%\, \bigg(\frac{3 \rm TeV}{M}\bigg)^2\bigg(\frac{E_B/M_N}{0.05}\bigg)\bigg( \frac{\langle \sigma_D v\rangle}{ \alpha/M_N^2}\bigg) \bigg(\frac{g_\star}{106.75}\bigg)^{1/2} \bigg(\frac{25}{z_f}\bigg)\,.
\ee

Contrary to the SM the production of deuterium is far from saturation for heavy dark matter. For this reason the actual abundance depends on the precise value of the cross-section that we will compute in section \ref{sec:calculation}.

\subsection{Neutral Constituents}\label{utterly}
When the fundamental fermions are SM singlets, the hadrons and mesons of the dark sector can be much lighter than the electroweak or even QCD scale, without occurring in constraints from LHC direct searches. As a consequence DM numerical densities comparable or even larger than visible matter (see eq.~\eqref{YDM}) become possible. Naively this is the most favourable situation for nucleosynthesis and even very large nuclei could be formed as advocated in \cite{Hardy:2014mqa}.  

Nevertheless in this section we would like to argue that, in presence of SM singlets, nucleosynthesis is very unlikely to take place unless  extra light degrees of freedom such as a dark photon \cite{1406.1171,McDermott:2017vyk} or a scalar  \cite{Wise:2014jva,Wise:2014ola,Gresham:2017zqi,Gresham:2018anj} are included (see also realisations in scenarios of mirror world \cite{Chacko:2018vss}). 
In absence of light fields external to the strong sector, such as a dark photon or light pions (see the discussion in the previous section), the first step of nucleosynthesis cannot occur as a $2\to 2$ process. 
The dark deuterium production must necessarily proceed through $3\to 2$ processes involving only baryon states, such as $N+N+N \leftrightarrow D + N$ reactions. Even at temperatures below $E_B$, when the production reaction is the only one that can occur, the fusion is suppressed as compared to the previous case by an additional power of $Y_{\rm DM}$. 
The Boltzmann equation in this case takes the form
\begin{equation}\label{boltzmann32}
\dot{n}_{\rm D}+ 3 H n_{\rm D}=  \langle \sigma_{3\to 2} v^2\rangle \left(n_N^3-  \frac {(n_N^{\rm eq})^2}{n_D^{\rm eq}}n_D n_N\right)\,.
\end{equation}
where we have introduced a generalised $3\to 2 $ cross-section with mass dimension $-5$. The thermally averaged $3\to 2$ cross-section is defined as
\begin{equation}
\langle \sigma_{3 \to 2} v^2 \rangle = \frac{n_D^{\rm eq}}{(n_N^{\rm eq})^2}  \langle\sigma_{2\to 3} v^2 \rangle  =\int \frac {d^3 p_1}{(2\pi)^3 2E_1}\frac {d^3 p_2}{(2\pi)^3 2E_2}\frac {d^3 p_3}{(2\pi)^3 2E_3} e^{-\frac {E_1+E_2+E_2}T} \left| {\cal M}_{NNN\to DN} \right|^2 
\end{equation}
For the analog of $s$-wave processes at low energy $\sigma v^2$ goes to a constant.
Following \cite{sigurdson-glueball} we estimate
\begin{equation}\label{stima3in2}
\langle\sigma_{3\to 2} v^2 \rangle\sim \frac {(4\pi)^3}{N^6} \frac 1 {M_N^5}\,.
\end{equation}

Introducing the deuterium baryonic fraction the Boltzmann equation  \eqref{boltzmann32} can be written as
\begin{equation}
\begin{split}
\frac {d X_{D}}{dz}&= a\frac{g_\star^{3/2} M_{\rm Pl}  E_{B}^4 \langle \sigma_{3\to 2} v^2\rangle}{z^5} Y_{\rm DM}^2 \left[(1-X_D)^3 -  \beta \big(\frac{g_N^2}{g_D g_\star}\big)  \big(\frac{z^{3/2} e^{-z}}{Y_{\rm DM}}\big)\big( \frac{M_N}{E_B}\big)^{3/2}\frac{X_D(1-X_D)}{2}\right]
\end{split}
\end{equation}
where $a=1.16$ is a numerical coefficient arising from the evaluation of $H$ and $s$. We can now compare the third line of the above equation with eq.~\eqref{boltzmann22} that sets the abundance of dark deuterium from $2\to 2$ processes. Compared to section \ref{subsec:SMcostituents}  for $z\gtrsim z_f$ the source term  decouples as fast as $z^{-4}$, is suppressed by higher powers of the binding energy and especially by one more power of $Y_{\rm DM}$. 

We then obtain the  estimate 
\begin{equation}
X_D\sim 10^{-13} \bigg(\frac{g_\star}{10}\bigg)^{3/2}\, \bigg(\frac{E_B/M_N}{0.01}\bigg)^4 \bigg(\frac{20}{z_f}\bigg)^4 \bigg(\frac{\langle \sigma_{3\to 2} v^2\rangle}{1/M_N^5}\bigg) \bigg(\frac{ \mathrm{GeV}}{M_N} \bigg)^3\,,
\end{equation}
which is utterly negligible in the relevant range of parameters.

\section{Production cross section of Dark Deuterium}\label{sec:calculation}

In this section we explain how the cross-section for formation of nuclei can be computed from first principles exploiting universal properties of short range nuclear interactions.
The main point is that, for shallow bound states such as nuclei where $E_B\ll M_N$, it possible to write a general effective theory of nucleons. This effective field theory (EFT) reproduces the 
effective range expansion of quantum mechanics and allows us to compute the properties of nuclei such as their production cross-section, see \cite{Kaplan:2005es} for a review. 
Here, we quickly outline the formalism and apply to the formation cross-section of dark deuterium.

\subsection{Dark Nucleon effective field theory}

The production of nuclei with $A=2$ is a process entirely analogous to the deuteron formation in the SM, $p n \to d \gamma$. This can be calculated in quantum mechanics with appropriate potentials, but, as noticed long ago \cite{Bethe:1949yr,Bethe:1950jm}, it does not depend much on the details of the potentials used, as long as they are short range. As emphasised in \cite{scaldeferri,Rupak:1999rk}, the generality of this phenomenon is immediately captured by the $\pi$-less effective theory of non-relativistic nucleons \cite{Kaplan:1998tg} that we briefly review. We refer the reader to the appendix and Refs. for more details.\footnote{In \cite{Braaten:2017gpq,Braaten:2017kci,Braaten:2017dwq} this formalism was applied to Wino scattering and annihilation.}

Since the energy scale relevant for nuclei formation is much below the pion mass it is useful to describe nucleons with a non relativistic lagrangian where the pions are integrated out, the $\pi$-less EFT \cite{Kaplan:1998tg}. Such theory is extremely simple because it only contains contact interactions among nucleons and couplings to SM gauge fields. Generalising the results of the Refs. above, the nucleons, in a generic isospin representation $\mathbf{r}$, are described by the effective lagrangian
\be\label{eff}
\mathscr{L}= N^\dag \left(i D_t + \frac{\vec D^2}{2M_N}+ \frac{ D^2_t}{2M_N}\right) N + \mathscr{L}_{4} + \frac{\kappa}{ M_N} g_2  N^\dag J^a (\vec\sigma \cdot \vec{B}_a) N\,,
\ee
where the covariant derivative $D_\mu= \partial_\mu - i g_2 A^a_\mu J^a$ contains the minimal coupling to SM gauge fields
and we have included 4-nucleons interaction and a magnetic dipole\footnote{In the case where the strong sector violates CP through a $\theta$ angle
an electric dipole is also present producing similar effects for deuterium formation. We will neglect this term.} interaction where $J^a$ is generator of SU(2)$_L$ in the nucleon rep. 
The coefficient $\kappa$ is expected to be of order unity (in the SM the isovector nuclear magnetic moment is $\kappa_V=2.35$).

At sufficiently low energies the leading interactions in $\mathscr{L}_{4}$ include only operators without derivatives, that can be decomposed into spin and isospin channels as
\be\label{L4}
\begin{split}
\mathscr{L}_{4}&= - \sum_{\mathbf{r},S}\frac{C_{\mathbf{r},S}}{4}\, (N [\CG^M_{\mathbf{r}}\otimes P^i_{S}] N)^\dag\, \, (N [\CG^M_{\mathbf{r}}\otimes P_{S}^i] N)\,,
\end{split}
\ee
where the matrices $\CG_{\mathbf{r}}$ and $P_S$  act on the isospin and spin space respectively.\footnote{For SU(2) the matrices $\CG$ can be identified with the Clebsch-Gordan coefficients, while the explicit expression of the spin projectors onto the spin 0 and 1 states are
\be
P_0=\frac{\sigma_2}{\sqrt{2}},\quad  P^i_1=\frac{\sigma_2\sigma_i}{\sqrt{2}}\,.
\ee
The labels $\mathbf{r}$ and $S$ identify the SU(2)$_L$ and spin representations, while $M$ and $i$ are the indices of such representations. }  

Remarkably, the lagrangian above also describes the non-perturbative bound state allowing to compute for example the production cross-section. A quick way to derive the main result is the following.
The non-relativistic amplitude for the elastic scattering of two nucleon has the general form in each isospin/spin channel $\mathbf{r},S$,
\begin{equation}\label{scattering-amplitude}
{\cal A}_{\mathbf{r},S} = \frac {4\pi } {M_N} \frac 1 {p \cot \delta_{\mathbf{r},S}- i p}
\end{equation}
where $\delta_{\mathbf{r},S}$ is the phase shift and $p=\sqrt{E M_N}$ is the nucleon momentum in the center of mass frame. 
For $s-$wave scattering in the low velocity regime one can show that $p \cot \delta_{\mathbf{r},S} = -1/a_{\mathbf{r},S} +O(p^2)$, where $a_{\mathbf{r},S} $ is the scattering length. 
This is know as the effective range expansion. When this is large and positive it follows that the amplitude has a pole at negative energy. 
From this we can recover the general relation between the scattering length of the binding energy of shallow bound states,
\begin{equation}
\frac{1}{a_{\mathbf{r},S}}\approx  \sqrt{M_N E_B}\,,
\end{equation}
where $E_B$ is the binding energy of lightest $s$-wave bound state. 

The coefficients of the 4-Fermi interactions in eq.~(\ref{L4})  must be fixed to reproduce the effective range expansion of nucleon nucleon elastic scattering.
 As shown in the appendix, to leading order in the derivative expansion but to all order in the scattering length $a_{\mathbf{r},S}$, on finds
\begin{equation}
C_{\mathbf{r},S}= \frac {4\pi}{M_N} a_{\mathbf{r},S}\,.
\end{equation}
Once this matching has been performed, the above lagrangian can be used to compute other processes, 
such as the production of deuterium (see \cite{scaldeferri} and refs for the SM case).

Indeed, the amplitude above determines also the coupling of two nucleons and deuterium as
\begin{equation}
g_{NND}^2 = {\rm Res}_{E=-E_B} \left[{\cal A}\right]= \frac {8 \pi  \gamma}{M^2}\,
\end{equation} 
where $\gamma=1/a$. This effective coupling can then be used to compute the interaction between 2-nucleons and the deuterium. 

Amazingly, we just need to study the elastic nucleon-nucleon ($NN$) scattering to infer all the quantities needed to perform the leading order calculation of the deuteron formation rate. Since this process occurs cosmologically at energy much below the mass of the pions (see also the discussion in section \ref{sec:properties}), it is very reasonable to use an effective field theory of non-relativistic nucleons (with SM quantum numbers) without the pions.

\subsection{Magnetic and Electric transitions}

The effective field theory in eq.~\eqref{eff} allows us to  compute the short distance cross section for the process $N+N\to D + V^a$ in terms of the binding energies and scattering lengths alone. 

The nucleus can be formed through emission of a SM gauge boson either through the electric coupling (minimal interaction) or through the magnetic dipole interaction in eq.~\eqref{eff}. For $\kappa\sim 1$, as expected for strongly coupled baryonic states, the two processes can have similar size. Importantly, different selection rules apply to electric and magnetic transition so that $\Delta L=1$ for the first and $\Delta S=1$ for the second. This implies a different velocity scaling of the cross-sections.

The amplitude for the formation of bound state can be simply computed with Feynman diagrams of the non-relativistic effective theory \eqref{eff} using eq. \eqref{eff} for the overlap of final state with the deuteron.\footnote{For Majorana bound states a factor $\sqrt{2}$ must be included in the amplitude to account for the normalisation of the wave-function of bound states made of identical particles.}
\be
\begin{tikzpicture}[line width=1.5 pt, scale=1.7]
	\draw[] (-1,0.5)--(0,0);
	\draw[] (-1,-0.5)--(0,0);
	\draw[vector,color= blue] (-0.5,0.25)--(0,.7);
	\node at (0.1,0) {$\boldsymbol\redcross$};
	\node at (-2,0) {$\mathcal{A}=$};
	\node at (-1,0) {$(\mathbf{r},s)$};
	\node at (.7,0) {$(\mathbf{r}',s')$};
	\node at (0.2,0.67) {$\varepsilon_{\lambda_a}$};
\end{tikzpicture}
\ee
The only subtlety arise for large scattering length of the initial state where an extra long distance contribution must be taken into account enhancing the tree level magnetic cross-section. This effect is discussed in detail in appendix \ref{sec:appA}.
The amplitudes for bound formation can be conveniently decomposed in the basis  of total spin and isospin of initial and final states $(NN, D)$ using the projectors of eq.~\eqref{L4}.
In the limit $v_{\rm rel}\ll 1$ one finds~\footnote{We neglect here the effect of non-abelian interactions \cite{Mitridate:2017izz}. This is justified if the nuclei are dominated by the strong interactions so that their size is smaller than the Coulombian Bohr radius $a_0^{-1}=\lambda \alpha_2 M_N/2$.}

\begin{eqnarray}
\mathcal{A}_{\rm mag}((NN)_{M,i}\to D_{M',i'} +V^a)&=& \frac{2 g_2\kappa}{M_N |\vec k|} g_{NND_{\mathbf{r}'}}(1- a_{\mathbf{r}} \gamma_{\mathbf{r}'})(\vec k \times  \vec \varepsilon_{(\lambda_a)})^{i+i'}\,\, C_{\cal J}^{a M M'}\,,\\
\mathcal{A}_{\rm ele}((NN)_{M,i}\to D_{M',i'} +V^a)&=&\frac{2 g_2}{M_N |\vec k|}g_{NND_{\mathbf{r}'}} \vec p \cdot \vec\varepsilon_{(\lambda_a)}\,\, \delta_{i i'}   C_{\cal J}^{a M M'}\,.
\label{amplitudes}
\end{eqnarray}
$M,M'$ are the indices of the isospin representations, while $i,i'$ are the indices of the total spin representations of initial and final states, and so these expressions should be read taking into account the selection rule of the magnetic and electric transition.
The group theory factor is \cite{Mitridate:2017izz}
\begin{equation}
C_{\cal J}^{a M M'}= \frac 1 2 {\rm Tr}[\CG_{\mathbf{r}'}^{M'} \{ \CG_{\mathbf{r}}^{M},J^a \}].
\end{equation}
The formulae above are general, and can be even applied to other gauge groups and different representations for the constituents of the bound state with minor modifications. For any more details we refer the reader to the appendix \ref{sec:appA}.

With our normalisations, the cross-section for the bound state formation can be computed with
\be\label{cross}
\sigma v_{\rm rel}  = \frac{|\vec{k}|}{8\pi^2}\int d\Omega_k |\mathcal{A}|^2\,,\quad \quad E_{\vec k}= E_B+ \frac{M_N}4 v_{\rm rel}^2\,.
\ee
This gives the following magnetic and electric cross-sections:

\paragraph{Magnetic cross-section}~\\
The averaged cross-section for the production of an $s$-wave bound state with energy $E_B$ and isospin quantum number $(I',M')$ through the emission of a (massive) vector boson $V^a$ from an initial state with $(I,M)$ at low velocity is given by,
\begin{equation}
(\sigma v_{\rm rel})_{aMM'}^{\rm mag}=\kappa^2\frac {2^8}{g_N^2} \sigma_0    \sqrt{1- \frac{M_a^2}{E_B^2}}\left(\frac {E_B} {M_N}\right)^{\frac 3 2}(1-a_{\mathbf{r}}\gamma_{\mathbf{r}'})^2  |C_{\cal J}^{a M M'}|^2\,,\quad\quad \sigma_0\equiv\frac{\pi \alpha}{M_N^2}\,.
\label{xsec:magnetic}
\end{equation}
Here $g_N=2(4) d_R$  is the number of degrees of freedom of the nucleon initial state for Majorana (Dirac) particles.
If the initial state supports a weakly bound state with $a_i=\sqrt{E_{B_i} M}$ otherwise $a_i$ is negative and 
can be large. For example in the SM the second term is $a_i \gamma_f \approx 5$ so it dominates \cite{scaldeferri,Rupak:1999rk}. 

At low velocities $\sigma v_{\rm rel}$ is constant as this correspond to a an $s$-wave capture with $\Delta L=0$ and $\Delta S=1$. The presence of a coherent contribution from the initial state scattering length can be understood by noticing that being a $s$-wave process, the initial state can be in principle have an unnaturally large coefficient in eq.~\eqref{eff} that need to be kept into account to all orders.

\paragraph{Electric cross-section (as in dipole approximation). }~\\
The averaged cross-section for the formation of an s-wave shallow bound state through dipole emission of a photon is found to be, 
\begin{equation}
(\sigma v_{\rm rel})_{aMM'}^{\rm ele}=\frac {2 S+1}{g_N^2}  \frac {2^6}3 \sigma_0 v_{\rm rel}^2 \sqrt{1- \frac{M_a^2}{E_B^2}} \sqrt{\frac {M_N} {E_B}} \left(1+\frac{M_a^2}{2E_B^2}\right)|C_{\cal J}^{a M M'}|^2
\label{xsec:electric}
\end{equation}
The velocity suppression follows from the fact that in dipole approximation $\Delta L=1$ so that an $s$-wave bound state is produce from a $p$-wave.
Note that the formula above differs by a numerical factor from the cross-section for the formation of Coulombian bound state \cite{Mitridate:2017izz}.
This is because the energy levels of $1/r$ potentials cannot be treated as shallow bound states.

From the formulae in each isospin channel for electric and magnetic transitions we can recover the component cross-section, relevant for example for indirect detection in \ref{sec:ID}
using the appropriate Clebsch-Gordan coefficients. Let us note that the formulae above are actually general and also apply to bound state made of different representations
and different global symmetry group.

\subsection{SE(s)}
\label{sec:SE}

The discussion so far has neglected the long distance Sommerfeld effect due to electro-weak interactions on the initial state. 
In a given SU(2)$_L$ channel, the cross section is multiplied by the corresponding factor depending on whether we deal with $s$ or $p$-wave initial states, see for example \cite{Iengo:2009ni}.
For massless mediators one finds,
\begin{eqnarray}\label{sommerfeld}
\SE_{s-\mathrm{wave}}&=&\frac {2\pi \alpha_{\rm eff}/v_{\rm rel}}{1-e^{-2\pi \alpha_{\rm eff}/v_{\rm rel}}}\approx \frac {2\pi\alpha_{\rm eff}}{v_{\rm rel}}\,,\\
\SE_{p-\mathrm{wave}}&=&\left[1+\left(\frac {\alpha_{\rm eff}}{v_{\rm rel}}\right)^2\right]\frac {2\pi \alpha_{\rm eff}/v_{\rm rel}}{1-e^{-2\pi \alpha_{\rm eff}/v_{\rm rel}}}\approx 2\pi \left(\frac {\alpha_{\rm eff}}{v_{\rm rel}}\right)^3\,.
\end{eqnarray}
Here $\alpha_{\rm eff}$ is the effective strength of the electro-weak forces in a given channel. Importantly, taking into account this effect, both the magnetic and electric transition rates have the same scaling with velocity, $\sigma v_{\rm rel}\propto 1/v$. Note that such enhancement of electric transitions is not present for deuteron in the SM so that the magnetic transition dominates at very low velocities. This can be different for dark nuclei. 

The approximate scalings above are accurate for $v_{\rm rel} \lesssim \alpha$. The effectiveness of the SE, 
then crucially depends on  the typical velocity during the nucleosynthesis. Dark deuterium forms at $T\sim E_B/z_f$ where $z_f\sim 20$, which implies $v_{\rm rel}\sim \sqrt{E_B/M_N}/5$ at that time.
Numerically the enhancement is more pronounced when the interaction are SU(2)$_L$ symmetric, as in this case the relevant coupling is $\alpha_2$, instead of $\alpha_{\rm em}$. 

Let us now discuss domain of validity of the massless approximations.
From the point of view of the initial state particles, the mediator masses can be neglected as long 
as the de-Broglie wave-length is smaller than the range of the electro-weak interactions $M_W^{-1}$. 
Therefore Sommerfeld effects are maximal if the above two conditions are met
\be
\frac{M_W}{M_N}\lesssim v_{\rm rel}\lesssim \alpha\,.
\ee
For $v_{\rm rel} \lesssim M_W/M_N$ the vector boson masses must be taken into account.
These has 2 effects, the first to freeze the enhancement to a constant value corresponding to the critical velocity,
\begin{equation}
\SE\rightarrow {\rm max}\bigg[2\pi \alpha_{\rm eff}\frac {M_N}{M_W},\,1 \bigg]\,
\end{equation}
This approximation gives results comparable to the analytic formulas derived using the Hulthen potential \cite{Mitridate:2017izz}.
In addition the mediator mass produces peaks in the cross-section. As well known the resonant behaviour originates from bound states of zero energy 
supported by the potential at the critical mass. Around the peak the SE has the model independent form \cite{Blum:2016nrz},
\begin{equation}
\SE \sim \frac{V_0}{M_N v_{\rm rel}^2/4+ |E_B|}
\label{SEpeak}
\end{equation}
where $V_0$ is the typical energy  and $E_B$ the energy of the bound state close to threshold in the initial state.

We also note that in the limit of large scattering length of the initial state the second term in (\ref{amplitudes}) can be interpreted as SE due to the strong interactions.
Indeed this is large when the initial state supports a bound state with energy $E_B\ll M_N$.
Using $1/a= \sqrt{E_B M_N}$ and extending the computation of bound state formation to finite velocity of nucleons 
one can check that the formula for bound state production through magnetic interaction ($s$-wave) 
reduces to eq. (\ref{SEpeak}) where $E_B$ is the energy of the bound state in the initial state while $V_0$ is the binding energy of the produced bound state.
The subleading terms in eq.~(\ref{xsec:magnetic}) is associated to the failure of factorisation between short and long distance effects.

\section{A case study: $\mathrm{SU(2)}_L$ triplet}\label{sec:triplet}

In this section we compute explicitly the production of deuterium in the $V$ model where the constituents are triplet of SU(2)$_L$. 
In this scenario, in the limit of vanishing SM couplings, the models enjoys an SU(3)$_F$ flavour symmetry and the lightest baryon multiplet is an octet. 
This decomposes under SU(2)$_L$ as
\be
\mathbf{8}= \mathbf{3}_{=V} +\mathbf{5}\,,
\ee
where with abuse of notation we named the triplet nucleon $V$. SM gauge interactions split the two multiplets as in eq. (\ref{nucleonsplitting}).
The triplet is expected to be the lightest state in light of the smaller weak charge. Since at the temperatures relevant for bound state formation the 
abundance of the quintuplet is exponentially suppressed we can focus on the nucleon $V$.

\begin{table}
\begin{center}
\begin{tabular}{cccc|c}
\hbox{Name}& $\mathbf{r}$ & $S$    &$\lambda$   & \hbox{Constituents}\\  \hline
$D_1$ &$1$ &  0&2 &  $V V$  \\ 
$D_3$ & 3 & 1  &1&     $V V$ \\ 
$D_5$ & 5 & 0  & -1 &    $V V$  \\  \hline
$T_1$ &1 & 3/2  &  2 &  $V D_3$ \\
$T_3^a$ & 3 & 1/2  &  0 &    $V D_1$ \\
$T_3^b$ & 3 &  1/2 & 1  &    $V D_3$\\
$T_3^c$ & 3 & 1/2  & 3 &    $V D_5$\\
$T_5^a$ & 5 & 1/2  & -1 &    $V D_3$  \\
$T_5^b$ & 5 & 1/2  & 1 &    $V D_5$ \\
$T_7$ & 7 & 1/2  &  -2 &   $V D_5$
\end{tabular}
\quad \quad \quad
\begin{tabular}{c|c|c|c}
$\mathbf{r}\leftrightarrow  \mathbf{r'}$ & $\sum\limits_{aMM'}|C_{\mathcal{J}}^{aMM'}|^2$  & $C_{\mathcal{J}}^{300}$ & $C_{\mathcal{J}}^{+01}$\\ [1.6ex]
\hline 
$1\leftrightarrow 3$  & $2$                      & $\sqrt{2/3}$  & $\sqrt{2/3}$           \\
$3\leftrightarrow 5$  & 5/2           & $\sqrt{1/3}$     &     $-\sqrt{1/12}$ 
\end{tabular}
\caption{ On the left quantum numbers of nuclear bound states (deuterium and tritium) for nucleons SU(2)$_L$ triplets. On the right group theory factors for the transition from an initial state with isospin $\mathbf{r}$ to a bound state with isospin $\mathbf{r'}$.}
\label{table:boundstates}
\end{center}
\end{table}

In absence of electro-weak interaction he nuclei with baryon number 2 belong to the product of two octet $\mathbf{8}\times \mathbf{8}=\mathbf{1}+ \mathbf{8}_S+ \mathbf{8}_A+ \mathbf{10} + \overline{\mathbf{10}}+\mathbf{27}_S$, where $S(A)$ refers to the symmetry of the isospin wave-function. 
Lattice studies indicate that all these representation actually form bound states \cite{Beane:2012vq} and also provide the binding energies in different channels. 
As expected in the flavor symmetric limit the singlet (corresponding to the $H-$baryon in QCD)
is the most bound. Following the discussion of section \ref{sec:properties}, we can work in a limit where we can classify the bound states according to SU(2)$_L$ representations, while neglecting the baryon in the \textbf{5} whose abundance is Boltzmann suppressed.
Therefore the dark deuterium of this model belongs just to the product
\be
V \times V = \mathbf{1}_S + \mathbf{3}_A + \mathbf{5}_S\,.
\ee
Anti-symmetry of the full wave-function implies that singlet and quintuplet of SU(2)$_L$ ($D_1$ and $D_5$) are spin-0 while isospin triplet are spin-1 ($D_\mathbf{3}$) (for s-wave bound states).
The triplet and quintuplet deuterons are branches of $\mathbf{8}_A$ and $\mathbf{8}_S$ respectively so they are heavier than the singlet, with splitting dominated by the strong interactions.

The classification can be generalised to larger nuclei.
The dark tritium made of 3 triplets has quantum numbers,
\begin{equation}
(D_\mathbf{1} + D_\mathbf{3} + D_\mathbf{5}) \times V = \mathbf{1}_S + 3 \times \mathbf{3}_A + 2 \times \mathbf{5}_S + \mathbf{7}
\end{equation}
where the symmetry of the wave-function determines the spin. We can estimate the electro-weak binding energy 
by adding a nucleon to the deuterium and taking into account the reduced mass.

We summarise the bound states up to dark Tritium in table \ref{table:boundstates}.

\begin{figure}[t]
\centering
\includegraphics[width=.45\textwidth]{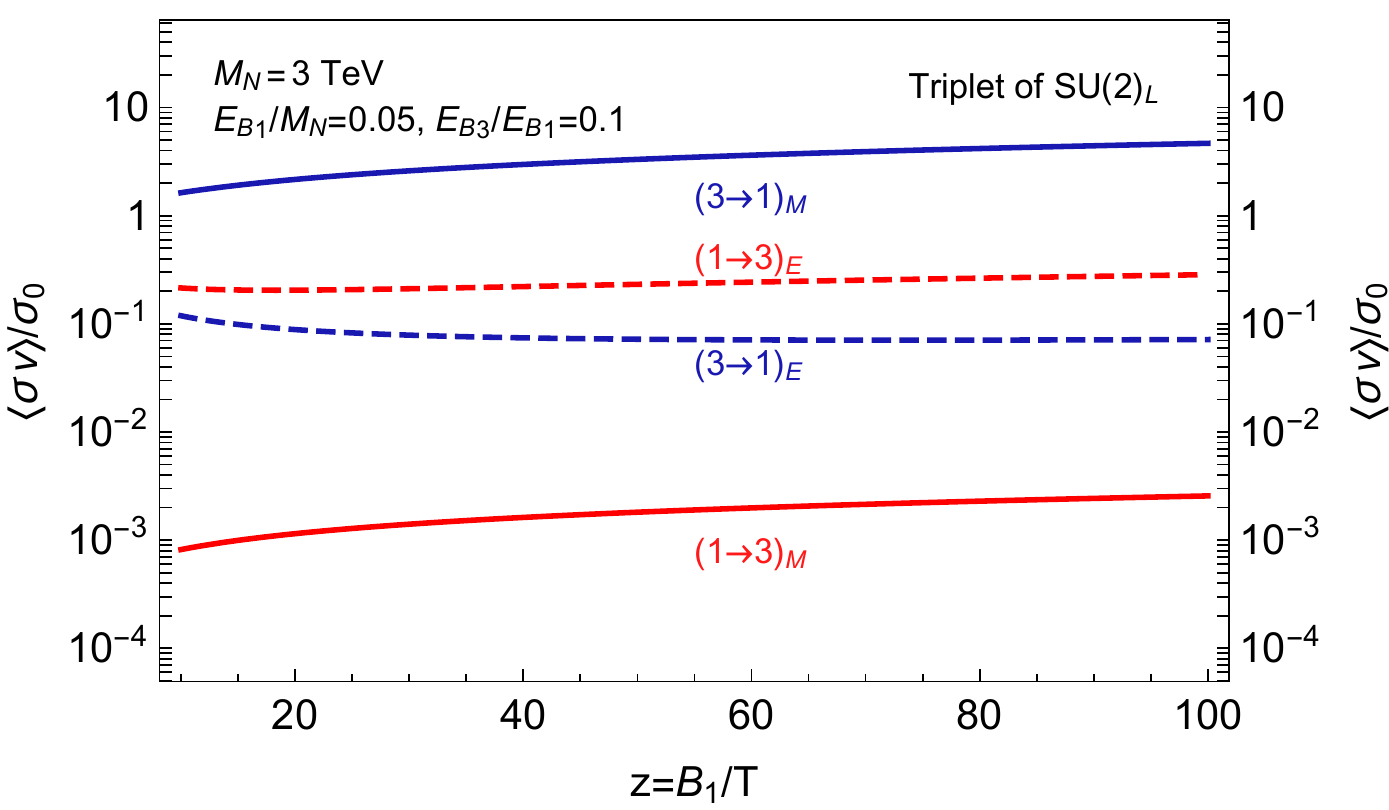}~~
\includegraphics[width=.45\textwidth]{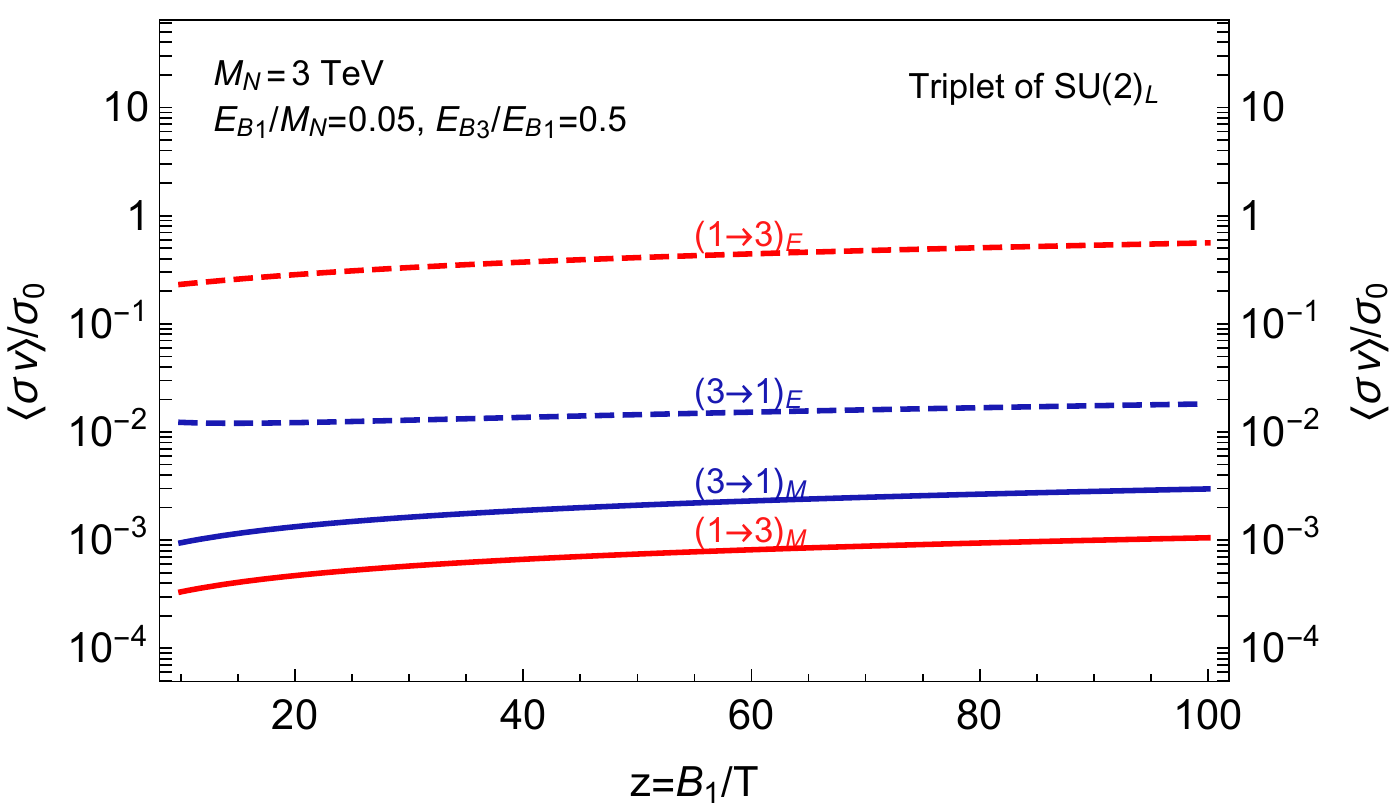}~~
\caption{\label{fig:xsec} Thermally averaged cross sections as a function of the temperature $z=E_B/T$. The rates are decomposed per channel and per type (magnetic $M$, and electric $E$).}
\end{figure}

\subsection{Relic abundances of deuterium made of $\mathrm{SU(2)}_L$ triplets}
To compute the abundance of deuterium we assume that the initial state is SU(2)$_L$ symmetric, i.e. we neglect the mass difference between $V_\pm$ and $V_0.$
Using \eqref{xsec:magnetic} and \eqref{xsec:electric} the cross-section for deuterium formation through emission of $W, Z, \gamma$, averaged over initial states, is approximately given by,
\begin{equation}
\begin{split}
(\sigma v_{\rm rel})_{3\to1}^a&= \sigma_a  \times \frac{2^7}{27}\sqrt{1- \frac{M_a^2}{E_{B_\mathbf{1}}^2}}\left[ \SE_3^{s}\,\kappa^2\left(\frac{E_{B_\mathbf{1}}}{M_N}\right)^{\frac 3 2}(1-a_{\mathbf{3}} \gamma_\mathbf{1})^2+ \SE_3^{p}\, \frac 1 {12}  \sqrt{\frac {M_N} {E_{B_\mathbf{1}}}}\left(1+ \frac {M_a^2}{2 E_{B_\mathbf{1}}^2}\right) v_{\rm rel}^2 \right]\\
(\sigma v_{\rm rel})_{1\to 3}^a&=\sigma_a  \times \frac{2^7}{27}\sqrt{1- \frac{M_a^2}{E_{B_\mathbf{3}}^2}}\left[ \SE_1^{s}\,\kappa^2\left(\frac{E_{B_\mathbf{3}}}{M_N}\right)^{\frac 3 2}(1-a_{\mathbf{1}} \gamma_\mathbf{3})^2+\frac 1 4 \SE_1^{p}\,   \sqrt{\frac {M_N} {E_{B_\mathbf{3}}}} \left(1+ \frac {M_a^2}{2 E_{B_\mathbf{3}}^2}\right) v_{\rm rel}^2 \right]\\
(\sigma v_{\rm rel})_{5\to 3}^a&=\sigma_a  \times \frac{2^5 \cdot 5}{27}\sqrt{1- \frac{M_a^2}{E_{B_\mathbf{3}}^2}}\left[ \SE_5^{s}\,\kappa^2\left(\frac{E_{B_\mathbf{3}}}{M_N}\right)^{\frac 3 2}(1-a_{\mathbf{5}} \gamma_\mathbf{3})^2+ \SE_5^{p}\,   \frac 1 4\sqrt{\frac {M_N} {E_{B_\mathbf{3}}}} \left(1+ \frac {M_a^2}{2 E_{B_\mathbf{3}}^2}\right)v_{\rm rel}^2 \right]\\
(\sigma v_{\rm rel})_{3\to 5}^a&=\sigma_a  \times \frac{2^5\cdot 5}{27}\sqrt{1- \frac{M_a^2}{E_{B_\mathbf{5}}^2}}\left[ \SE_3^{s}\,\kappa^2\left(\frac{E_{B_\mathbf{5}}}{M_N}\right)^{\frac 3 2}(1-a_{\mathbf{3}} \gamma_\mathbf{5})^2+\SE_3^{p}\,   \frac 1 {12} \sqrt{\frac {M_N} {E_{B_\mathbf{5}}}} \left(1+ \frac {M_a^2}{2 E_{B_\mathbf{5}}^2}\right)v_{\rm rel}^2 \right]
\end{split}
\label{xsecSU2}
\end{equation}
where 
\begin{equation}
\sigma_a=\frac {\pi \alpha_a}{M_N^2}\,,\quad \quad \alpha_{W,Z,\gamma}= \alpha_2\times [2,\,c_W^2,\,s_W^2]
\end{equation}

The first term in bracket corresponds to magnetic $\Delta S=1$ transitions and the second the electric one $\Delta L=1$. 
The triplet can be produced either from a singlet or quintuplet channels. The scattering length are given by $1/a_i\approx \sqrt{M_N E_{B_i}}$.
If a state is unbound the formulas above still apply with a negative scattering length. This is the case of deuterium in the SM where $nn$ is weakly unbound 
producing a large scattering length. 

In Fig. \ref{fig:xsec} we show the numerical values of the electric and magnetic cross-sections $1\leftrightarrow 3$ normalised to $\sigma_0=\pi\alpha_2/M^2$ for various choices of the binding energies.
Due to the SE described in section \ref{sec:SE} the p-wave electric cross-section can be larger than the magnetic one.

To compute the abundance of deuterium in principle one should write a different Boltzmann equation for each bound state.
We can however simplify the problem by noting that transition between different 
bound states are fast so that they are in equilibrium among them (see section \ref{sec:darkforce} and appendix \ref{sec:appB}). This implies that $n_{D_i}/n_{D_j}= n_{D_i}^{\rm eq}/n_{D_j}^{\rm eq}$.
The abundance of deuterium is then determined by eq. (\ref{boltzmann22}) with the effective cross-section  and degrees of freedom,
\begin{equation}\label{effective-xsec}
(\sigma v_{\rm rel})^{\rm eff} = \sum_i (\sigma v_{\rm rel})_{i}\,,~~~~~~~~~~~g_D^{\rm eff}(T)= \sum_i g_{D^i} \exp{\left[- \frac{E_{B_1}-E_{B_i}}{T}\right]}\,.
\end{equation}

\begin{figure}[t]
\centering
\includegraphics[width=.45\textwidth]{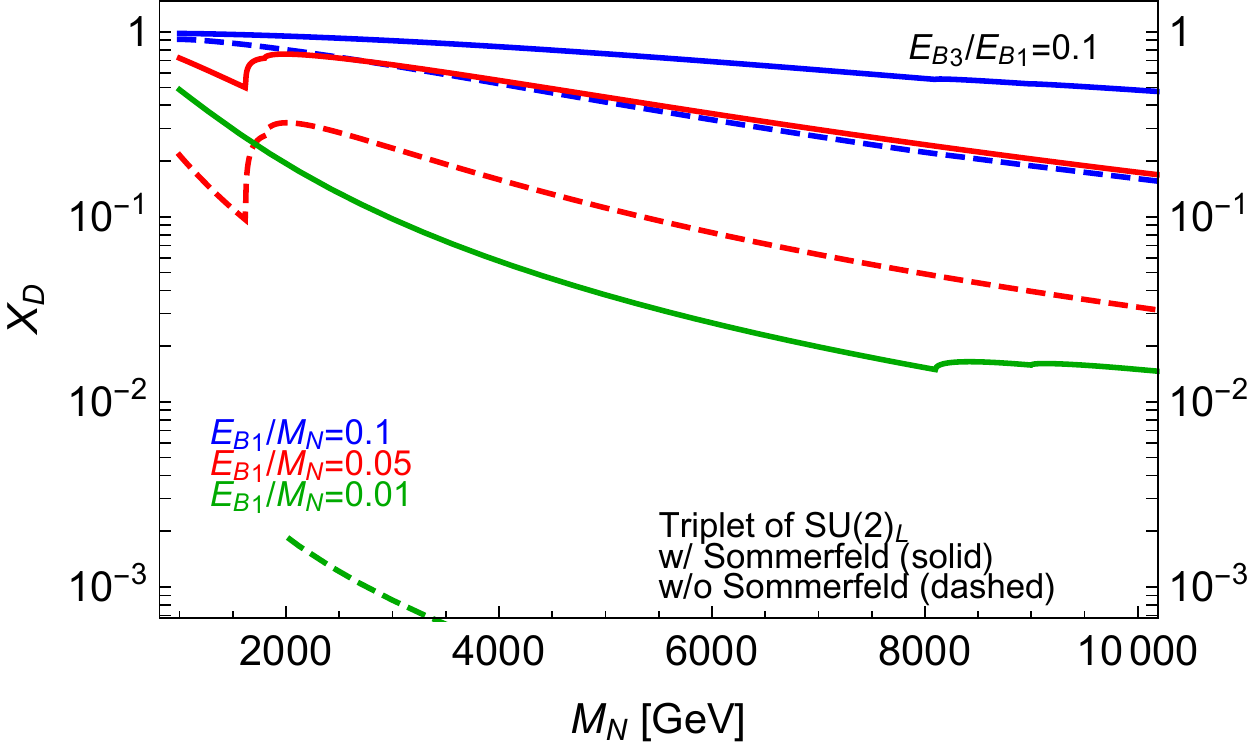}~
\includegraphics[width=.45\textwidth]{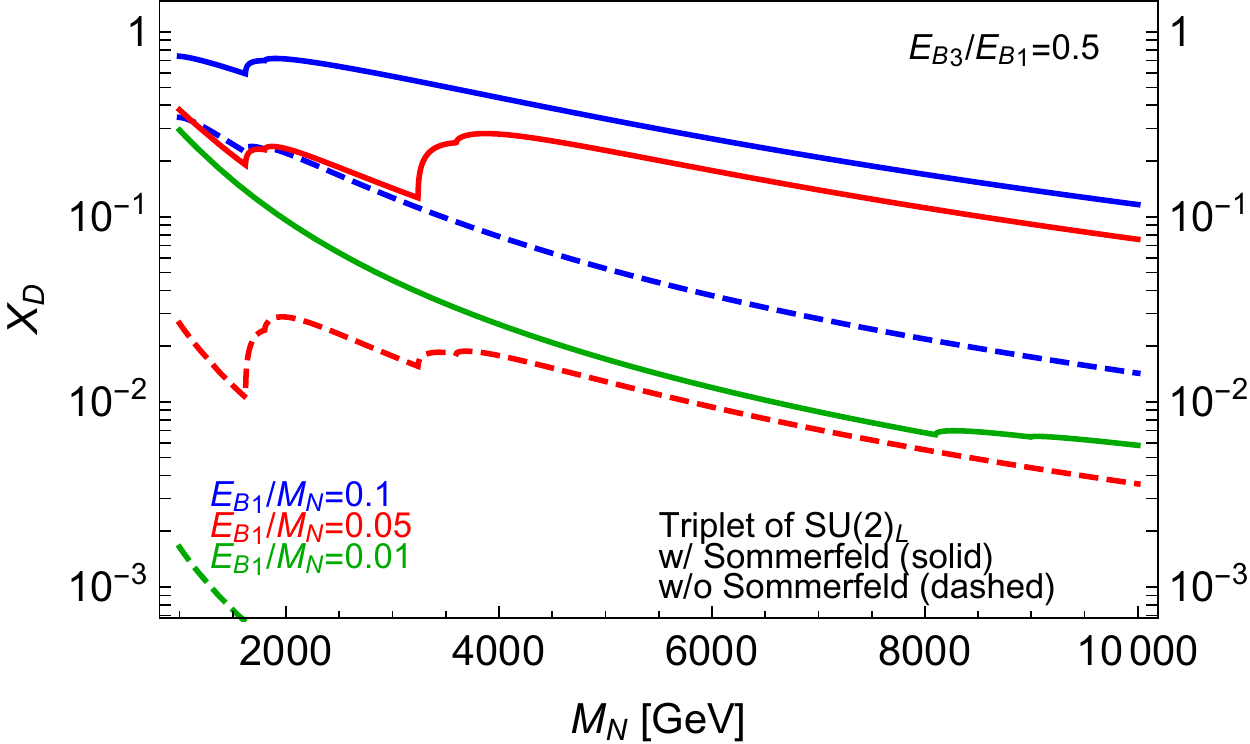}
\caption{\label{fig:dark-deuterium} Dark deuterium mass fraction $X_D$ as a function of the nucleon mass $M_N$ for several choices of the binding energies (with and without the Sommerfeld effect).}
\end{figure}

In figure \ref{fig:dark-deuterium} we present the total mass fraction of deuterium $X_D$ as a function of the $M_N$ for different choices of binding energies of the singlet ($D_\mathbf{1}$) and triplet ($D_\mathbf{3}$) deuterium. We assume for simplicity that $D_\mathbf{5}$ does not play a role even though being the least bound isotope with baryon number 2 it could enhance the production of $D_\mathbf{3}$ through a large scattering length. Solid lines correspond to the abundance including electro-weak SE effects while dashed lines are obtained with the short distance cross-sections. The SE enhancement has significant impact especially because it eliminates the velocity suppression of electric transitions. For $E_B< M_W$ only the photon can be emitted with smaller cross-section in light of the electric coupling and multiplicity. The transition between SU(2)$_L$ symmetric emission and photon emission originates the features in the plot.

Differently from the SM most of the baryons can form deuterium only for large binding energies. An order 1 fraction of deuterium can only be obtained for TeV masses, around the experimental collider bound. 
The large mass scale associated with DM with electroweak charges is the principal obstruction to convert into deuterium and then heavier nuclei an O(1) fraction of DM. Notice that there is no reduction in the mass fraction of heavy nuclei caused by the reduction of $V^\pm$ from the decay $V^\pm \to \pi^\pm V^0$, which happens only $T<100$ MeV. On the contrary in the SM the main limitation to form deuterium is the neutron decay.

\subsection{Production of nuclei with $A\geq 3$}
The formation of dark tritium $T$ is determined schematically by the following reactions
\begin{eqnarray}\label{chain}
D&:&\quad V+V \rightleftarrows D+W\,\\
T&:&\quad D+V \rightleftarrows T+W\,,\ D+D\rightleftarrows T+V\,,
\end{eqnarray}
where we allow for weak and strong $T$ formation processes, the latter without the emission of SM radiation. As for the deuterium, dark tritium is produced via exothermic reactions when $E_T-E_D>0$ (weak process) and/or  $E_T - 2E_D>0$ (strong process). If they hold, when the temperature drops below $E_T-E_D$ and $E_T-2E_D$ the dissociation of dark tritium is exponentially suppressed. The most favourable situation arises, as in the SM, for $E_T-2E_D > E_D$, such that at the time of deuterium formation, the dissociation of tritium is already ineffective.

These assumptions greatly simplify the discussion and they allow us to estimate the upper bound of dark tritium abundance. The Boltzmann equations including deuterium and tritium are given by
\begin{eqnarray}
\label{uno}\small
X_D'(z)&=& 2\frac{z_0}{z^2}\bigg[ (1-X_D-X_T)^2 - \beta \frac{(\frac{M_N}{E_D})^{3/2}\, z^{3/2}e^{-z}}{g_D^{\rm eff} Y_{\rm DM}} \frac{X_D}{2}- \frac{b_1}{2} (1-X_D-X_T) X_D - \frac{b_2}{2} X_D^2\bigg]\,,\\
\label{due}
X_T'(z)&=& 3\frac{z_0}{z^2}\bigg[ \frac{b_1}{2} (1-X_D-X_T) X_D + \frac{b_2}{4} X_D^2\bigg]\,.
\end{eqnarray}
Where we have introduced the following notation
\be
z_0\equiv  c\sqrt{g_*} M_{\rm Pl} E_{D} Y_{\rm DM} \langle \sigma_{D} v\rangle_{\rm eff}\,, \quad b_1\equiv \frac{ \langle \sigma_{T} v\rangle_{\rm eff}}{\ \langle \sigma_{D} v\rangle_{\rm eff}}\,,\quad b_2\equiv \frac{ \langle \sigma_{T} v\rangle^{\rm strong}_{\rm eff}}{\ \langle \sigma_{D} v\rangle_{\rm eff}}\,.
\ee
In deriving the above Boltzmann equations we have written effective production rates including the effect of nearby bound states with the same baryon number. In particular, $\langle \sigma_{D} v\rangle_{\rm eff}$ is defined in eq.~\eqref{effective-xsec} and the reactions for Tritium are defined similarly, although we would like to distinguish the weak $ \langle \sigma_{T} v\rangle_{\rm eff}$ from the strong $\langle \sigma_{T} v\rangle_{\rm eff}^{\rm strong}$ process. This inclusive approach allows us in principle to take into account all possible bound states of Table \ref{table:boundstates}. 

When the dissociation rate of deuterium, $D+W \to V+V$, becomes exponentially suppressed, we see that the terms proportional to $b_1$ and $b_2$ tend to transfer a fraction of  $X_D$ to $X_T$ with an overall decoupling as fast as $1/z^2$ (neglecting possible enhancements from Sommerfeld effect). As expected the strong fusion reactions ($b_2$-term) have smaller rates per deuteron since they are proportional to $X_D^2$, partially reducing the increase in the hard rate, $b_2/b_1\approx 1/\alpha$. 

Knowing the reaction rates one can simply solve numerically the above set of equations, however it is interesting to analyse it in the limit of small nuclei mass fractions (i.e. $X_D\ll 1$). For the most realistic cases we expect $z_0\approx O(1)$, therefore the non-trivial evolution happens when the overall rate is very small $z_0/z\ll 1$. By contrast, in the BBN we have $z_0\approx 10^4$, which would allow for a fast fusion of heavier nuclei if bottlenecks for the binding energy were absent \cite{ABG}. The formation of dark Tritium is then further suppressed and we have
\be\label{tritium}
X_D\approx 2\frac{z_0}{z_f} \,,\quad X_T\approx \frac{3}{8}X_D^2 b_1 +\frac{1}{8}X_D^3 b_2\,.
\ee
These expression are accurate as long as one can neglect the loss terms in the Boltzmann equation for $X_D$ and at leading order in $b_1$ and $b_2$. This is reliable as long as $X_D\ll \mathrm{min}[1/b_1,1/\sqrt{b_2}]$.

\subsection{Indirect Detection}\label{sec:ID}
The possibility of forming bound states is relevant for indirect detection as it would lead to the emission of monochromatic photons 
of energy equal to the binding energy of the nucleus. This possibility is of great interest also because asymmetric scenarios typically do not produce 
indirect detection signals since the DM cannot annihilate into SM particles. It is also independent on whether dark nuclei are synthesised cosmologically.
We leave a detailed study to  \cite{indirect} and here outline the main results.

In the $V$ model at late times DM is made of the neutral component $V_0$ of the nucleons plus a model dependent population of deuterium, also in the neutral component. 
We will neglect the deuterium population for the purpose of this section. If the nuclear binding energy of deuterium is larger than $M_W$ one can form the triplet deuterium 
via the tree-level process $V_0V_0\to D_3^\pm W^\mp$. The magnetic transition \eqref{xsec:magnetic} including long distance nuclear effects (see below) gives 
\begin{equation}
\big[\sigma_{V_0V_0\to D_3^\pm W^\mp} v_{\rm rel}\big]_{\rm hard}=\frac {\pi \alpha_2}{M_N^2}  \times \frac{2^8}{9}\sqrt{1- \frac{M_W^2}{E_{B_\mathbf{3}}^2}}\kappa^2\left(\frac{E_{B_\mathbf{3}}}{M_N}\right)^{\frac 3 2}\left[(1-a_{\mathbf{1}} \gamma_\mathbf{3})+\frac 1 2 (1-a_{\mathbf{5}} \gamma_\mathbf{3})   \right]^2\,,
\end{equation}  
where we neglected the electric transition which 
is velocity suppressed in absence of SE. For $E_{B_\mathbf{3}}<M_W$ the $W$ is off-shell leading to a suppressed cross-section.

Another contribution arises from the SE due to SU(2)$_L$ gauge interactions and nuclear forces.
Thanks to this effect, two neutral nucleons can form deuterium (in neutral component) through the emission of a photon. Physically this is possible because $| V_0 V_0 \rangle_S^\ell$ is not a mass eigenstate and can  oscillate into $| V_+ V_- \rangle_S^\ell$. In the symmetric limit the mixing can be extracted from the Clebsch-Gordan coefficients,
\be\label{CGcoeff}\small
| V_+ V_- \rangle=\frac {1}{\sqrt{3}} | 0 0 \rangle+\frac{1}{\sqrt{2}} | 1 0 \rangle +\frac {1}{\sqrt{6}} | 2 0 \rangle\,, \quad
| V_+ V_0 \rangle  =\frac{1}{\sqrt{2}} | 1 1 \rangle+\frac{1}{\sqrt{2}} | 2 1 \rangle \,,\quad 
| V_0 V_0 \rangle =-\frac{1}{\sqrt{3}} | 0 0 \rangle +\sqrt{\frac 2 3} | 2 0 \rangle\,.
\ee
The following processes are then possible
\begin{eqnarray}
|V_0 V_0\rangle_{S=0}^{s} &\to& |V_- V_+\rangle_{S=0}^s \ \to\ D_3^0 + \gamma\,,\\
|V_0 V_0\rangle_{S=1}^{p} &\to& |V_- V_+\rangle_{S=1}^p \ \to\ D_3^0 + \gamma\,,
\end{eqnarray}
such that we have formation of neutral $D_3$ from an initial state $|V_0 V_0\rangle_0$ in the singlet or quintuplet of SU(2)$_L$, either from $p$-wave spin-1 through an electric transition or from $s$-wave spin-0 via a magnetic transition.

\paragraph{Indirect signals from Sommerfeld of $\mathrm{SU(2)}_L$ gauge interactions}
At the low velocities relevant for indirect detection the electro-weak symmetry breaking effects can be important and a numerical solution is required, see \cite{Hisano:2006nn,Cirelli:2007xd}.
For $s$-wave the SE due to electro-weak interactions  is identical to the one of Wino DM $\chi$ studied widely studied in the literature, see \cite{Cohen:2013ama}. In particular we are interested in the long distance effects that allow the neutral Winos state $\chi^0\chi^0$ to oscillate into $\chi^+\chi^-$. We define as $\SE_{00\to +-}$ the corresponding Sommerfeld factor of the Wino that can be derived from the following potential 
\be
\label{eq:S0triplet}
V_{Q=0}^{S=0}=\bordermatrix{&+&0\cr -&2\Delta M-A&-\sqrt{2}B \cr 0& -\sqrt{2}B&0 }\,,
\ee
where $A =   \alpha_{\rm em}/r + \alpha_2 c_{\rm W}^2 e^{-M_Zr}/r$, 
$B=\alpha_2e^{-M_Wr}/r$
and $\Delta M$ is the mass splitting produced by electroweak symmetry breaking, equal to $\Delta M=165$ MeV. 

Noticing that in our case we have the same SE of Wino DM, so that we can  exploit the calculation performed for the Wino and apply it to the dark deuterium. Therefore, the indirect detection signal can be simply calculated as
\be\label{SEWino}
\sigma_{V_0V_0\to  D^0_{\mathbf{3}}+\gamma} = \SE_{00\to + -}\, \big[\sigma_{V_+V_-\to D^0_{\mathbf{3}}+\gamma}\big]_{\rm hard}\,.
\ee
The short-distance contribution to the cross-section of the charged components can be computed using \eqref{xsec:magnetic} and \eqref{CGcoeff},
\begin{equation}
\big[ \sigma_{V_+V_-\to D^0_{\mathbf{3}}+\gamma} v_{\rm rel}\big]_{\rm hard}= \kappa^2 \frac {2^7}{9}\frac{\pi \alpha_{\rm em}}{M_N^2}\left( \frac {E_{B_{\mathbf{3}}}}{M_N}\right)^{\frac 3 2}\left[(1-\sqrt{E_{B_{\mathbf{3}}}/E_{B_{\mathbf{1}}}}) +\frac 1 2(1-a_{\mathbf{5}} \gamma_\mathbf{3})\right]^2\,.
\end{equation}

The annihilation rate of Winos into photon pairs, $\sigma_{\chi^0\chi^0 \to \gamma\gamma +\frac12\gamma Z}$, has the property that both the $\gamma\gamma$ and $\gamma Z$ final states are reached from $\chi^0\chi^0$ with the same $\SE_{00\to +-}$. This allows us to provide an alternative form for eq.~\eqref{SEWino} in terms of Winos cross-sections \cite{Hisano:2006nn}

\be\label{SE00}
\SE_{00\to +-} =\frac{[\sigma_{\chi^0\chi^0 \to \gamma\gamma +\frac12\gamma Z} v_{\rm rel}]_{\rm full}}{\big[ \sigma_{\chi_+ \chi_- \to \gamma\gamma +\frac12\gamma Z} v_{\rm rel}\big]_{\rm hard}}\,,\quad  \quad [\sigma_{\chi_+ \chi_- \to \gamma\gamma +\frac12\gamma Z} v_{\rm rel}]_{\rm hard} =\frac {\pi \alpha_{\rm em}\alpha_2}{M_N^2}\,.
\ee

\begin{figure}[t]
\centering
\includegraphics[width=.45\textwidth]{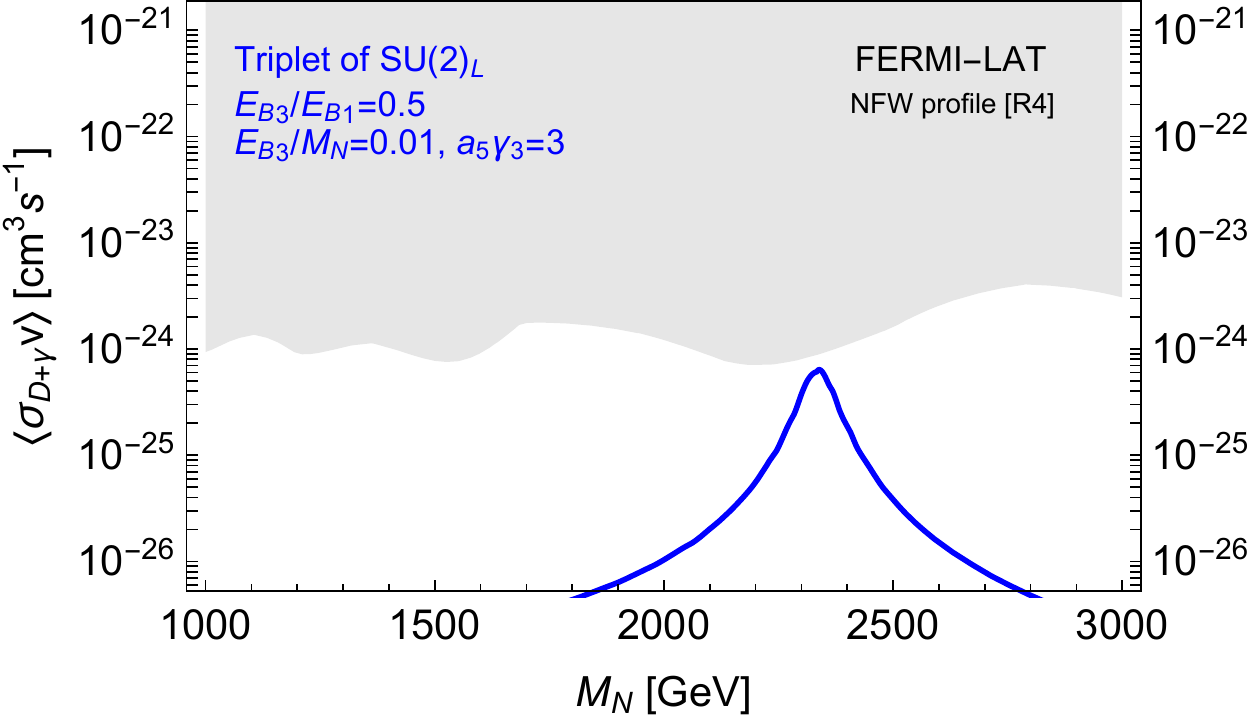}~
\includegraphics[width=.45\textwidth]{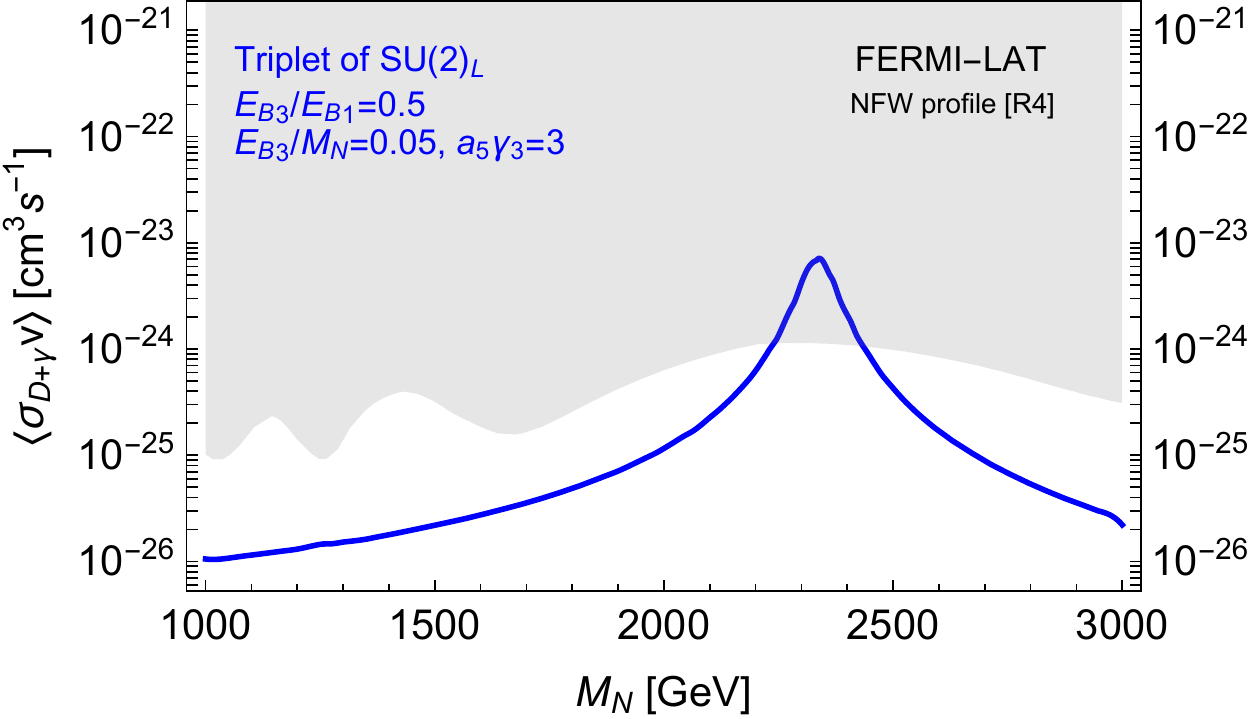}~
\caption{\label{fig:indirect} Indirect detection bound from emission of photon lines of energy $E_{D_3}$ through magnetic coupling ($\kappa=1$). 
Exclusion limits from FERMI are obtained by recasting the results in \cite{FERMI} to account for the different ratio of photon energy and DM mass.
SE for the triplet has been extracted from \cite{Cohen:2013ama}.}
\end{figure}

Due to SE the effective cross-section has peaks whose location is identical to the one of the Wino. The first peak appears for $M_N\approx 2.3$ TeV, 
but the energy of the emitted  photon is now equal to the binding energy of the bound states rather than DM mass. 
The sensitivity to these photons can be studied according to the recipes described in \cite{Cirelli:2018iax}.
In Fig. \ref{fig:indirect} we recast the bounds from FERMI from the galactic center \cite{FERMI}. 
Formation of deuterium through magnetic emission of photons  is strongly constrained for $E_B>M_N/10$.

\paragraph{Indirect signals from strong interactions}
Beside the standard SE  due to electro-weak interactions also the strong interactions can enhance the cross-section when the initial channel 
supports a bound state at threshold. This effect is captured by the  scattering length in eq. (\ref{xsec:magnetic}). We can compute the cross-section 
relevant for indirect detection in the SU(2)$_L$ symmetric limit. Since the quintuplet is expected to be less attractive or even weakly repulsive the largest effect will be associated to the quintuplet initial state.
Using  the Clebsch-Gordan coefficients of eq.~\eqref{CGcoeff} and neglecting further enhancement from electro-weak interactions  we find,
\begin{equation}
\sigma_{V_0 V_0 \to D_3^0 + \gamma}v_{\rm rel} \approx \kappa^2 \frac {2^7 }{9} \frac{\pi \alpha_{\rm em}}{M_N^2} \left(\frac{E_{B_\mathbf{3}}}{M_N}\right)^{\frac 3 2} \times  \frac {M_N E_{B_{\mathbf{3}}}}{1/a_5^2+M_N^2 v_{\rm rel}^2/4}\,.
\end{equation}

\section{Singlet Models}\label{sec:darkphoton}

\begin{figure}[t]
\centering
\includegraphics[width=.45\textwidth]{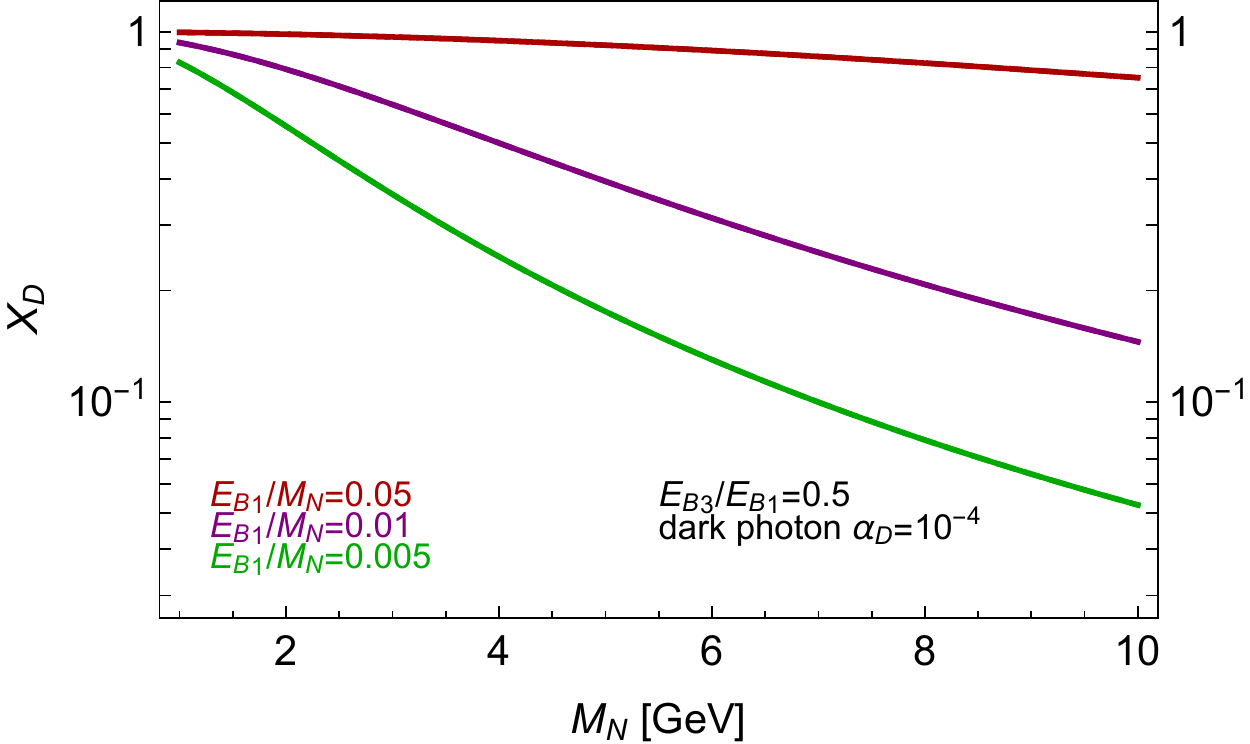}~
\includegraphics[width=.45\textwidth]{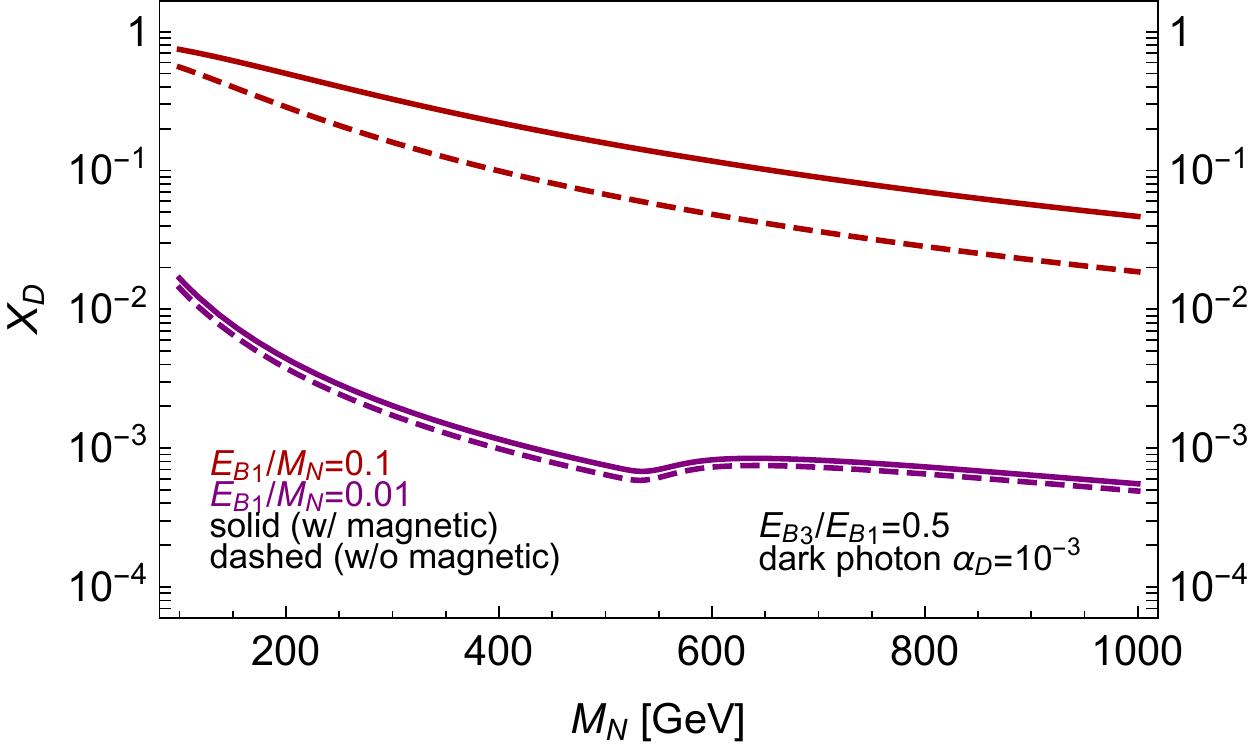}~
\caption{\label{fig:dark-photon} Dark deuterium abundance in models with 2 degenerate flavours and a dark photon. }
\end{figure}

In this section we briefly study bound state formation in a model with singlets. 
For simplicity we consider the minimal scenario with SU(3) gauge group and 2 degenerate flavours. 
In this case the strong dynamics is as in the SM and the lightest baryon is an SU(2)$_F$ doublet, $Q$. 
The dark deuterium is an $s$-wave bound state in the singlet or a triplet rep of SU(2)$_F$ with spin 1 or 0 respectively,
\be
Q \times Q = \mathbf{1}_1 + \mathbf{3}_0\,.
\ee
Strong interactions favour the SU(2)$_F$ singlet as the lightest state (the analog of the SM deuterium) that will then be absolutely stable. 
Differently from the SM also the triplet could be bound. We will assume that the temperature of the dark sector is of the order 
of the SM bath which can be  realised introducing heavy fermions  charged under the SM.

Singlets allow the mass scale to be much lower than TeV.
As discussed in section \ref{utterly} despite the larger numerical density of nucleons deuterium cannot be formed because $3\to 2$ processes are very suppressed.
This conclusion can be avoided introducing a dark photon that carries away the energy. Leaving the model building aspects to future work we assume the fermions to have opposite unit charges, $Q_D=2 J_3$ so that the nucleons have  charges $\pm 1$. This could be modified with different charge assignments with minor changes to the nuclei formation.

Neglecting model dependent SE due to the dark photon the cross-section for formation of deuterium through emission of a  dark photon is,
\begin{equation}
(\sigma v_{\rm rel})_{NN\to D_{\mathbf{1}(\mathbf{3})}+\gamma'}= \frac {\pi \alpha_D}{M_N^2}  \times 2^4\left[ \kappa^2\left(\frac{E_{B_{\mathbf{1}(\mathbf{3})}}}{M_N}\right)^{\frac 3 2}(1-a_{\mathbf{3}(\mathbf{1})} \gamma_{\mathbf{1}(\mathbf{3})})^2 +  \frac 1 {4(12)}  \sqrt{\frac {M_N} {E_{B_{\mathbf{1}(\mathbf{3})}}}} v_{\rm rel}^2 
\right] 
\label{xsecdarkphoton}
\end{equation}
where we used $\sum|C_{\mathcal{J}}^{3MM'}|^2=1/4$ for the doublet rep and assumed $m_{\gamma'}\ll E_B$.

In Fig. \ref{fig:dark-photon} we show the abundance of deuterium obtained integrating the Boltzmann equation and assuming that no other nuclei are produced.
For masses in the GeV range, suggested by the coincidence of DM and visible matter density, the production of deuterium is very efficient even for dark photon couplings as small as $10^{-4}$.
Note that electric transitions, even though velocity suppressed, are relevant for small binding energies as the one of deuterium in the SM ($E_B/M_N=0.0022$). Differently from the SM for degenerate masses we do not expect significant bottlenecks so that  production of heavy elements can  proceed unsuppressed. For  100 GeV masses the production of deuterium is again suppressed due to the smaller numerical density of baryons.

\section{Conclusions}\label{sec:conclusions}

Big Bang Nucleosynthesis is a cornerstone of the cosmological history of the Universe and one might  wonder if a similar process could take place in the dark sector.
In this work we have studied the synthesis of dark nuclei  in theories where DM is a baryon of a new gauge interaction. Such models, motivated by the idea 
of the accidental stability of DM, also predict the existence of stable and metastable nuclei with SM charges that could be formed during the evolution of the Universe. 

The key step for Dark Nucleosynthesis is the formation of dark deuterium, a nucleus with baryon number 2. This process requires energy to be released 
and it is automatically possible through emission of SM gauge bosons in models where the fundamental constituents have electro-weak charges.
Remarkably, the relevant cross-section can be determined from first principles in terms of the binding energies of nuclei. 
To this aim, we have found it useful to employ pion-less EFT for the nucleons. This construction reproduces the effective range expansion of quantum mechanics with 
short range potentials \cite{Kaplan:1998tg} and it allows us to compute in general the cross-sections for the production of  shallow bound states  expected for nuclear interactions.  
We note that the cross-sections differ from the ones often used in the literature, depending in a non-trivial way on binding energies and velocity.
In particular electric transitions grow for small binding energies while magnetic transitions decrease. Electric transitions moreover are velocity suppressed unless 
enhanced by Sommerfeld enhancement.

Having determined the relevant cross-sections one can solve the Boltzmann equations for the abundance of deuterium and heavier elements. 
In the case of DM with electro-weak charges, for example a baryon triplet of SU(2)$_L$ here studied in detail, 
one finds that only a fraction of DM binds into deuterium unless  binding energies much larger than in nuclear physics are assumed.
The main reason for this is simply the numerical density of DM: direct searches imply that the DM mass is greater than 1 TeV for electro-weak constituents. 
The small numerical abundance compared  to the SM nucleons suppresses the production of nuclei. Therefore dark BBN ends after deuterium formation leaving
a fraction of deuterium plus small traces of tritium. In minimal models with singlets, production of dark deuterium is kinematically forbidden and nucleosynthesis cannot start.
This conclusion changes completely including a dark photon: for DM masses in the GeV range, deuterium is formed very efficiently due to the large density and heavier nuclei 
can thus be formed through fusion reactions allowing to populate nuclei up to large atomic numbers. 

While in this work we have focused on asymmetric scenarios where DM mass is a free parameter, our results can be extended to symmetric models,
producing extra annihilation channels and nuclei-anti-nuclei. For the simplest scenarios however the critical abundance is reproduced for masses around 100 TeV 
and no significant fraction of nuclei is produced.

The formation of DM nuclei is interesting experimentally as it can lead to novel signatures in DM indirect detection, even in  asymmetric scenarios and 
change the prediction for direct detection experiments due to the different composition of DM. 
The emission of monochromatic photons with energy equal to the nucleus binding energy is a smoking gun of  dark nuclei that could be searched experimentally \cite{indirect}.
 
This work extends the computation of perturbative bound state formation in \cite{Mitridate:2017izz} to strongly coupled bound states.
We have provided general formulae, that could be used in other contexts, for electric and magnetic interactions also taking into account important 
long distance effects associated to bound states close to zero energy. It would be interesting to generalise this formalism to the fusion of strongly coupled 
bound states as well studying perturbative bound states within this framework.

{\small
\subsubsection*{Acknowledgements}
We wish to thank Hyung Do Kim, Gordan Krnjaic, Maria Paola Lombardo, Filippo Sala and Juri Smirnov for useful discussions. 
AT is partially supported by the grant ``STRONG" from the INFN.  We thank the Galileo Galilei Institute for Theoretical Physics for the hospitality during the completion of this work.
}

\appendix

\section{Details on Dark Deuterium formation}\label{sec:appA}
In this appendix we summarize the derivation of \cite{Kaplan:1998tg,Kaplan:1998we} that can be used to compute the dark deuterium formation directly from eq.~\eqref{eff}.

Since the coefficients $C_{\mathbf{r},S}$ of eq.~\eqref{L4} determine the scattering amplitude in each spin-isospin channels, those associated with shallow bound states are large and should be treated as a relevant coupling \cite{Kaplan:1998tg,Kaplan:1998we}. In an expansion in $p$ but at all order in the coupling $C_{\mathbf{r},S}$, the  amplitude for 2-nucleons elastic scattering can be written as
\be
	\begin{tikzpicture}[line width=1.2 pt, scale=1.4]
	\node at (-0.5,0) {$\displaystyle \mathcal{A}([NN\to NN]_{\mathbf{r},S})=\frac{-i C_{\mathbf{r},S} }{1+ iC_{\mathbf{r},S} \Sigma(p)} = $};
	\draw[line width=1.2pt] (2,0.5)--(2.5,0);
	\draw[line width=1.2pt] (2,-0.5)--(2.5,0);
	\draw[line width=1.2pt] (3,0.5)--(2.5,0);
	\draw[line width=1.2pt] (3,-0.5)--(2.5,0);
	\draw[fill] (2.5,0) circle (.1cm);
	\node at (3.5,0) {$=$};
	\draw[line width=1.2pt] (4,0.5)--(4.5,0);
	\draw[line width=1.2pt] (4,-0.5)--(4.5,0);
	\draw[line width=1.2pt] (5,0.5)--(4.5,0);
	\draw[line width=1.2pt] (5,-0.5)--(4.5,0);
	\node at (4.5,-0.5) {$C_{\mathbf{r},S}$};
	\node at (5.5,0) {$+$};
	\draw[line width=1.2pt] (6,0.5)--(6.5,0);
	\draw[line width=1.2pt] (6,-0.5)--(6.5,0);
	\node at (6.5,-0.5) {$C_{\mathbf{r},S}$};
	\draw[line width=1.2pt] (8,0.5)--(7.5,0);
	\draw[line width=1.2pt] (8,-0.5)--(7.5,0);
	\node at (7.5,-0.5) {$C_{\mathbf{r},S}$};
	\draw[] (7,0) circle (.5cm);
	\node at (7,0) {$\Sigma$};
	\node at (8.5,0) {$+\,\cdots$,};
	\end{tikzpicture}
	\ee
which should  be matched with eq.~\eqref{scattering-amplitude}. Notice that this calculation requires a renormalisation of the theory, since $\Sigma(p)$ is divergent. 
A useful scheme is the on-shell momentum subtraction (or the analogous PDS \cite{Kaplan:1996xu} scheme), which amounts to say that the 1-loop term plus the counter-term evaluated at $p=i\mu$ in dimensional regularisation should be equal to $C_{\mathbf{r},s}(\mu)$. In this scheme
\be
\Sigma(p)=\int \frac{d^4q}{(2\pi)^4}\frac{i}{(q^0+E-\vec{q}^{\, 2}/2M_N)}\frac{i}{(-q^0 -\vec{q}^{\, 2}/2M_N)}=-i\frac{M_N}{4\pi}(\mu + i p)\,, \quad p=\sqrt{M_N E}\,.
\ee
To leading order in the momentum expansion eq.~\eqref{scattering-amplitude} is reproduced with,
\be
C_{\mathbf{r},S}(\mu)=\frac{4\pi a_{\mathbf{r},S}}{M} \frac{1}{1-\mu a_{\mathbf{r},S}}\,.
\ee
When $a_{\mathbf{r},S}$ is large and positive it signals the presence of a bound state in the corresponding channel. 
The scattering length is related to the binding energy, since the pole in the amplitude corresponds to a momentum $p=i\sqrt{M_N E_B}$, giving
 $a_{\mathbf{r},S}=1/\sqrt{M_N E_B}\equiv 1/\gamma_{\mathbf{r}}$. Alternatively, one can solve for $C_{\mathbf{r},S}$ requiring a pole in the amplitude for a given binding energy. 

Expanding the amplitude at the pole we can also determine the interaction of the bound state $D_{\mathbf{r},S}$ with the two nucleons. Namely the coupling is the square 
root of residue of the amplitude at the pole. Therefore quite generally we can write as effective coupling

\be\label{coupling}
\begin{tikzpicture}[line width=1.2 pt, scale=1.6]
	\node at (-4,0) {$\displaystyle g_{NND_{\mathbf{r}}}= \frac{\sqrt{8\pi \gamma_{\textbf{r}}}}{M_N}, \quad \quad \gamma_{\mathbf{r}}=\sqrt{B_{\mathbf{r}} M_N}$};
	\draw[line width=1.2pt] (-1,0.5)--(0,0);
	\draw[line width=1.2pt] (-1,-0.5)--(0,0);
	\node at (0.1,0) {$\otimes$};
	\node at (0.4,0) {$\mathbf{r},S$};
	\node at (1,0) {$=$};
	\node at (-1,0) {$g_{NND}$};
	\node at (1.5,0) {$\bigg($};
	\node at (1.8,0) {$\mathrm{Res}$};
	\draw[line width=1.2pt] (2,0.5)--(2.5,0);
	\draw[line width=1.2pt] (2,-0.5)--(2.5,0);
	\draw[line width=1.2pt] (3,0.5)--(2.5,0);
	\draw[line width=1.2pt] (3,-0.5)--(2.5,0);
	\node at (3.5,0) {$\bigg)^{\frac{1}{2}}$};
	\node at (3.1,-0.1) {\tiny$E=-B_{\mathbf{r}}$};
	\draw[fill] (2.5,0) circle (.05cm);
	\node at (-1,0) {$g_{NND}$};
	\end{tikzpicture}\,.
\ee
Once the coupling of the bound state to two nucleons is determined, 
one can compute with ordinary Feynman diagrams the amplitude relevant for dark deuterium formation. This formalism is particularly powerful as it allows to 
compute systematically the effects to higher orders and allows to resum  effects associated to large scattering lenghts. 

In this work we are interested in all the processes of the type
\be
\mathcal A ( \big[N+N\big]^{M,i}_{\mathbf{r}, s} \to D^{M',i'}_{\mathbf{r}', s'} + V^a_\lambda)\,.
\ee
where a bound state is formed through emission of a gauge boson. This implies the existence of selection rules: 
In the explicit case of  SU(2)$_L$, all the leading order amplitudes correspond to a $\Delta I=1$ transition. Instead, for spin and angular momentum there are two possibilities: $i)$ a magnetic dipole interaction of eq.~\eqref{eff} which corresponds to a $\Delta S=1$ transition; $ii)$ an electric transition (in dipole approximation) with $\Delta L=1$ that originates from the covariant derivative in eq.~\eqref{eff}.

\subsection{Details on the magnetic transition}
In the SM the deuteroun is an isospin 0 and spin 1 state. At low energy the formation is dominated by the magnetic transition) from an initial spin singlet isospin triplet channel \cite{Rupak:1999rk}. The selection rule implies that it proceeds from an $s$-wave initial state ($^1S_0$) so that $\sigma v$ goes to a constant for slow nucleons. Moreover this process is significantly enhanced  by the large scattering length of the $^1S_0$ channel. 

Our scenario differs from the SM for the different group theory structure.  To leading order for large scattering lengths of the initial state
there are two diagrams that contribute,
\be\label{fig:magnetic}
\begin{tikzpicture}[line width=1.5 pt, scale=1.7]
	\draw[] (-1,0.5)--(0,0);
	\draw[] (-1,-0.5)--(0,0);
	\draw[vector,color= blue] (-0.5,0.25)--(0,.7);
	\node at (0.1,0) {$\boldsymbol\redcross$};
	\node at (-2,0) {$\mathcal{A}_1=$};
	\node at (-1,0) {$(\mathbf{r},S)$};
	\node at (.7,0) {$(\mathbf{r}',S')$};
\end{tikzpicture}~~~~
\begin{tikzpicture}[line width=1.5 pt, scale=1.7]
	\node at (-2,0) {$\mathcal{A}_2=$};
	\draw[] (-1,0.5)--(0,0);
	\draw[] (-1,-0.5)--(0,0);
	\draw[vector,color= blue] (0.5,0.5)--(0.7,.7);
	\draw (0.5,0) circle (.5cm);
	\draw[fill] (0,0) circle (.1cm);
     \node at (1.1,0) {$\boldsymbol\redcross$};
	\node at (-1,0) {$(\mathbf{r},S)$};
	\node at (2,0) {$(\mathbf{r}',S')$};
	\node at (0.5,0) {$\Pi$};
\end{tikzpicture}
\ee

The crossed circle represents the deuteron  coupling \eqref{coupling}, while the filled circle corresponds to the insertion of the resummed amplitude for the scattering in the $(\mathbf{r} ,S)$ channel. 
The loop integral appearing in the amplitude $\mathcal{A}_2$ is finite giving,
\be
\begin{split}
\Pi(k) &= \int \frac{d^4q}{(2\pi)^4}\frac{i}{-q^0-\vec{q}^{\, 2}/(2M_N)}\frac{i}{q^0-\vec{q}^{\, 2}/(2M_N)}\frac{i}{q^0-|k|-\vec{q}^{\, 2}/(2M_N)}\\
&=\int \frac{d^3 q}{(2\pi)^3}\frac{M_N^2}{ q^4+M_N |\vec k|  q^2 }=\frac{M_N^2}{4\pi \sqrt{M_N|\vec k|}}+O(E^2)\,.
\end{split}
\ee

The ratio of the two amplitudes for small energy is
\be
\frac{\mathcal{A}_2}{\mathcal{A}_1}= - a_{\mathbf{r}} \gamma_{\mathbf{r}'}\,,
\ee
where we have used the fact that $k=E_{B_{\mathbf{r}'}}$. By virtue of this relation, the leading order term of the magnetic transition can be computed easily just focussing on $\mathcal A_1$.
Notably, $\mathcal{A}_2$ can be of the same order and might dominate over the first one.

In full generality, to leading order in the momentum expansion,  the magnetic amplitude from an initial state $\big[N+N\big]_{\mathbf{r}, s}$ to a final state with $D_{\mathbf{r}',s'} + W^a_\lambda$ is given 
\be\label{eq:generic-magnetic-transition}
\mathcal A ( \big[N+N\big]^{M,i}_{\mathbf{r}, s} \to D^{M',i'}_{\mathbf{r}', s'} + W^a_\lambda) =  \frac{2g_2\kappa}{M_N|\vec k|} g_{NND_{\mathbf{r}'}}(1- a_{\mathbf{r}} \gamma_{\mathbf{r}'})(\vec k \times  \vec \varepsilon_{(\lambda_a)})^{i+i'}\,\, \delta_{s+s',1} C_{\cal J}^{a M M'}
\ee
which is valid for any group and any representation.  For non zero velocity of the nucleons a similar formula can be derived.

\section{Bound State decays}
\label{sec:appB}
The formalism above also allows to compute decay rates.
Focusing on  bound states with $\ell=0$ and spin $0,1$ the decay of excited states to lower levels occurs at leading order  in $\alpha$  
via the magnetic interaction. The decay rate is given by
\be
\Gamma = \frac{|\vec k|}{2\pi}|\mathcal{A}_{D_i \to D_j W}|^2\,.
\ee
The amplitude for the transition between two bound states with $\Delta I=1$ and $\Delta S=1$, with the emission of a gauge boson, is given by the 1-loop graph in eq.~\eqref{fig:magnetic} evaluated at the difference of binding energies between the two states,
\be
\begin{tikzpicture}[line width=1.5 pt, scale=1.7]
	\node at (-1,0) {$\mathcal{A}=$};
	\draw (0.5,0) circle (.5cm);
	\node at (-0.1,0) {$\boldsymbol\redcross$};
       	\node at (1.1,0) {$\boldsymbol\redcross$};
	\node at (-0.5,0) {$(\mathbf{r},s)$};
	\node at (1.5,0) {$(\mathbf{r}',s')$};
	\node at (0.5,0) {$\Pi$};
	\node at (1.53,0.7) {$\vec\varepsilon_{\lambda_a}$};
	\draw[vector,color= blue] (0.5,0.5)--(1.2,.7);
	\node at (5,0) {$\displaystyle=\frac{2g_2\kappa}{M_N}g_{NND_{\mathbf{r}'}}g_{NND_{{\mathbf{r}}}} \Pi(|\vec k|)|\vec k| (\vec k \times \vec\varepsilon_{\lambda})^{i+i'}\delta_{s+s',1} \, C_{\cal J}^{a M M'}$,};
\end{tikzpicture}
\ee
By energy conservation $|\vec k |= E_B$, which gives $\Pi=M_N^2/(4\pi \sqrt{M_N  E_B})$. Summing over photon polarisations and spin indices, one finds
\be
\Gamma =\kappa^2 \frac{(256\alpha_2)}{g_{D_\mathbf{r}}}   \frac{(E_{B_{\mathbf{r}}}-E_{B_{\mathbf{r}'}})^2}{M_N^2}\sqrt{E_{B_{\mathbf{r}}} E_{B_{\mathbf{r}'}}}\,\sum_{a M M'} |C_{\cal J}^{a M M'}|^2\,.
\ee

\pagestyle{plain}
\bibliographystyle{jhep}
\small
\bibliography{biblio}

\end{document}